\documentclass[aps,prb,twocolumn,showpacs,preprintnumbers,superscriptaddress]{revtex4}
\usepackage{amsmath}
\usepackage{amssymb}
\usepackage{graphicx}% Include figure files
\usepackage{dcolumn}% Align table columns on decimal point
\usepackage{bm}% bold math
\usepackage{color}% bold math
\usepackage{tikz}
\usetikzlibrary{shapes.geometric}
\usepackage{pinlabel}
\usepackage{ulem}
\newcommand{\drop}[1]{}

\definecolor{NavyBlue}{rgb}{0.00,0.00,0.90}
\definecolor{Red}{rgb}{1.00,0.00,0.00}
\definecolor{green}{rgb}{0.18,0.55,0.34}
\definecolor{Magenta}{rgb}{1.00,0.00,1.00}
\definecolor{electricblue}{rgb}{0.49, 0.98, 1.0}
\definecolor{Orange}{rgb}{1.00,0.50,0.00}
\definecolor{violet}{rgb}{0.56, 0.0, 1.0}

\begin{document}

\title{Electronic and optical properties of doped TiO$_2$ by many-body perturbation theory}
      
%% AUTHORS
%%
%\author{Michael O. \surname{Atambo}}
\author{Michael O. Atambo}
%\email[corresponding author: ]{michaelontita.atambo@unimore.it}
\affiliation{Centro S3, CNR--Istituto Nanoscienze, 41125 Modena, Italy}
\affiliation{Department of Physics, Mathematics, and Informatics, University of Modena and Reggio Emilia, 41125 Modena, Italy}
\author{Daniele Varsano}
\email{daniele.varsano@nano.cnr.it}
\affiliation{Centro S3, CNR--Istituto Nanoscienze, 41125 Modena, Italy}
\author{Andrea Ferretti}
\affiliation{Centro S3, CNR--Istituto Nanoscienze, 41125 Modena, Italy}
\author{S. Samaneh Ataei}
\affiliation{Centro S3, CNR--Istituto Nanoscienze, 41125 Modena, Italy}
\author{Marilia J. Caldas}
\affiliation{Instituto de F\'\i{}sica, Universidade de S{\~a}o Paulo,
  Cidade Universit\'aria, 05508-900 S{\~a}o Paulo, Brazil}
\author{Elisa Molinari}
\affiliation{Centro S3, CNR--Istituto Nanoscienze, 41125 Modena, Italy}
\affiliation{Department of Physics, Mathematics, and Informatics, University of Modena and Reggio Emilia, 41125 Modena, Italy}
\author{Annabella Selloni}
\affiliation{Department of Chemistry, Princeton University, Princeton, NJ}

%
% PACS
%
%\pacs{}
\date{\today}

%
% ABSTRACT
%
%\input{abstract}

\begin{abstract}

Doping is one of the most common strategies for improving the photocatalytic and solar energy conversion properties of TiO$_2$, hence an accurate theoretical description of the electronic and optical properties of doped TiO$_2$ is of both scientific and practical interest.
In this work we use many-body perturbation theory techniques to investigate two typical n-type
dopants, Niobium  and Hydrogen,
in TiO$_2$ rutile.
%Niobium as aprototype substitutional defect ( TiO$_2$:Nb$_{\textrm{Ti}}$) and  Hydrogen, as an interstitial dopant (TiO$_2$:H$_{i}$) 
Using the GW approximation to
determine band edges and defect energy levels, and the Bethe Salpeter equation for the calculation of the absorption spectra, 
%we show that many body corrections are crucial for accurately  describing the energy position of 
%the impurity  states and the band gap narrowing of the doped material.
we find that the
defect energy levels form non-dispersive bands %associated with localized states 
lying $\simeq 2.2~eV$  above the top of the 
corresponding valence bands
($\simeq  0.9~eV$ below the conduction bands of the {\it pristine} material). 
The defect states are also responsible for the appearance of low energy absorption peaks that enhance the solar spectrum absorption of rutile.  
%Interestingly, 
The spatial distributions of the
excitonic wavefunctions  associated with these low energy excitations are very different for the two dopants,
%we find a very different nature of the excitonic wavefunctions in terms of localization, 
suggesting a larger mobility of photoexcited electrons in 
Nb-TiO$_2$.

\end{abstract}

%
%%%%%%%%%%%%%%%%%%%%%%%%%
% Main body
%%%%%%%%%%%%%%%%%%%%%%%%%
%
\maketitle

%\input{body_resub}

%
% paper content
%

%%%%%%%%%%%%%%%%%%%%%%
%\section*{TO-DO LIST}
%%%%%%%%%%%%%%%%%%%%%%

%%%%%%%%%%%%%%%%%%%%%
\section{Introduction}
%%%%%%%%%%%%%%%%%%%%%
%
Titanium dioxide (TiO$_2$) is 
widely used in
%one of the most abundant, chemically stable and functionally versatile oxide materials, with applications ranging from white pigment in paints to
photocatalysis and solar energy conversion,
but its efficiency is limited by the
large band gap ($\sim$ 3.0 and 3.2 eV for rutile and anatase, respectively) that severely reduces the photoabsorption of visible light~\cite{gratzel2001photoelectrochemical,diebold2003surface}.
To improve the conductivity and photocatalytic properties, doping is often employed,
and a large variety of dopants, both metals and non-metals, have been explored.~\cite{asahi_visible-light_2001,di_valentin_characterization_2005} 
%
%For example, the substitution of a lattice Oxygen with a non-metal atom like Nitrogen leads to acceptor states above the valence band maximum (VBM),~\cite{di_valentin_characterization_2005} whereas Oxygen substitution with Fluorine gives rise to donor states below the conduction band minimum (CBm).~\cite{di_valentin_theory_2005}
%
Numerous computational studies of doped TiO$_2$ have  also been reported.~\cite{di_valentin_reduced_2009,di2013trends} 
In this context, studies including an
accurate theoretical description of the band gap and impurity levels combined with a detailed analysis of the influence of doping on the optical absorption can contribute new insights of scientific and practical interest as well.

Two of the most common dopants in TiO$_2$ are Niobium and atomic Hydrogen. 
They both act as donors, with Nb substituting a Ti atom, and Hydrogen forming an interstitial defect bound to an Oxygen of the lattice.~\cite{di_valentin_reduced_2009}
Nb is usually described as a shallow donor that greatly improves the electrical conductivity of TiO$_{2}$.~\cite{doi:10.1021/acs.chemmater.6b02031} At the same time, however, valence-band photoemission spectra of Nb-doped rutile show a peak at about 0.8-1.0 eV below the Fermi energy, indicating an electron in a localized Ti$^{3+}$ state similar to that observed in Oxygen deficient rutile.~\cite{morris2000photoemission} 
This apparent contradiction between ultraviolet photoelectron spectroscopy (UPS) and electrical measurements can be explained within a small polaron model, where the vertical excitation energy of the polaron, as observed in UPS, can be much larger than the energy of adiabatic excitations involved in electrical measurements.~\cite{di_valentin_reduced_2009} 
It is generally agreed that the polaron model 
applies to most defects and impurities in rutile, 
including interstitial Hydrogen.~ \cite{stavola2014contrasting,sezen2014probing,hupfer2017photoinduced}
In particular, recent infrared absorption measurements in reduced and hydrogenated rutile have been interpreted on the basis of hybrid density functional calculations, where the self-trapping energy of a conduction band electron to form a localized Ti$^{3+}$ ion was predicted to be $\sim 0.3$ eV.~\cite{sezen2014probing}

The electronic structure of doped TiO$_2$ has been the subject of many theoretical studies. 
It is now widely accepted that while standard local and semi-local Density Functional Theory (DFT) do not adequately describe the polaronic character of defect and impurity states in TiO$_2$, DFT+U and hybrid functionals are effective in determining the structure and energetics of the polaronic states. However, also these approaches are insufficient for precisely describing the electronic energy levels. For instance, hybrid functionals often  overestimate the TiO$_2$ band gap thus leading to an uncertainty in the position of the impurity levels.~\cite{gerosa2017accuracy}
Recent many-body perturbation theory (MBPT) studies\cite{thulin_calculations_2008,chiodo_tailoring_2010,kang_quasiparticle_2010,alves-santos_electronic_2014} of the band structure and absorption spectrum of pristine TiO$_2$  have in fact provided evidence that approaches beyond DFT are required for a quantitative description of photoemission and light absorption experiments in this material. 

The aim of this work is to investigate the electronic structure and optical properties of Nb- and H-doped TiO$_2$, using a rigorous approach that goes beyond mean-field theory. We adopt MBPT, namely the GW approximation for the quasi-particle energies and the Bethe Salpeter equation for the optical spectra.  Following a description of the adopted methodology and computational details in Sec.~\ref{sec:methods},  our results for the quasi-particle electronic structure and optical properties are presented and discussed in Sec.~\ref{sec:results}. Conclusions are presented in Sec.~\ref{sec:conclusions}. We find that both Nb and H doping introduce optically allowed transitions well below the intrinsic optical gap of rutile, thus improving its ability to absorb the solar energy.

%==========================================================================================
%%%%%%%%%%%%%%%%%%%%%
\section{Methods}
%%%%%%%%%%%%%%%%%%%%%
\label{sec:methods}

\subsection{Ground state calculations}

Ground state calculations were performed using Kohn-Sham (KS) DFT~\cite{kohn-sham65pr}  within the plane wave-pseudopotential scheme, as implemented in the Quantum ESPRESSO package.~\cite{giannozzi_quantum_2009,giannozzi_advanced_2017} 
We used the gradient-corrected
GGA-PBE~\cite{perd+96prl} exchange-correlation functional and norm-conserving
ONCV pseudopotentials,~\cite{hamann_optimized_2013} including the semi-core states for Titanium (3s and 3p).  We obtained converged KS eigenstates and eigenvalues using a wavefunction kinetic energy cut-off of 80 Ry.

Doped rutile was modelled using a 2$\times$2$\times$3 supercell (SC) containing 72 atoms (73 atoms in the case of H doping),  including one Nb replacing a Ti atom (one interstitial H impurity), which corresponds to a dopant concentration of $\sim 4\%$ (referred to the number of Ti ions). With this SC, reciprocal space was sampled  
using a Monkhorst-Pack grid of 2$\times$2$\times$2. 
For the sake of comparison, pristine rutile was modelled using both the same SC and the primitive cell with an equivalent $\mathbf{k}$-point sampling. 
We verified that the two descriptions provide equivalent results, and therefore in the following only the results obtained  using the  2$\times$2$\times$3 supercell are reported. 

%\subsection{Structural Properties}
Relaxed geometries for both pristine and defective TiO$_2$ were computed including a Hubbard-U
term~\cite{coco-degi05prb} in the Ti 3d orbitals.
This scheme has been shown to work well 
for the description of polaronic effects in TiO$_2$, particularly for U values in the range 3-4 eV.~\cite{finazzi_excess_2008,morgan_dft+u_2007,PhysRevB.78.241201,setvin2014direct}
Here we use U = 3.5 eV, a value close to that given by linear response theory,~\cite{PhysRevB.78.241201}
that has been extensively tested and validated in previous studies~\cite{zhao2013effective,dette_tio2_2014,selcuk_structural_2018,aschauer2012hydrogen} 
With this choice, the computed cell parameters of pristine TiO$_2$ rutile are $a=4.64 $\AA{} and $c=2.96$ \AA, in good agreement with the experimental values, $a=4.59$ \AA{} and $c=2.95$ \AA.\cite{doi:10.1021/ja01623a004}

%\subsection{Electronic Structure}
\subsection{MBPT calculations}

The electronic and optical properties of pristine and doped rutile TiO$_2$ were investigated
using the GW method for the self energy operator $\Sigma$ and the Bethe-Salpeter equation (BSE) for the absorption spectra.\cite{onid+02rmp,Hanke74,Hanke80,Strinati82,Strinati88}
Specifically, we used the single-shot G$_0$W$_0$ approximation by treating the frequency dependence of the dielectric matrix via the Godby-Needs (GN) plasmon-pole model (PPM).\cite{godb-need89prl}

All the MBPT calculations were carried out using the plane-wave code Yambo.\cite{mari+09cpc,sang+18tobe}
In the G$_0$W$_0$ calculations, converged results were obtained by considering
1200 bands (corresponding to 59.53 eV above the energy of the top of the valence band), and a kinetic energy cutoff of 8 Ry for the screening dielectric matrix. GW corrections were introduced on top of DFT-PBE calculations, without the inclusion of Hubbard U.
The Bethe-Salpeter equation was built considering vertical transitions involving 20 occupied and 40 empty bands.

In the single shot G$_0$W$_0$ approximation, G$_0$
is built from single particle KS orbitals,
and the quasi-particle energies are obtained as a
first order perturbation correction. 
In practice, quasi-particle (QP) energies are obtained as:
   \begin{equation}
       E^{QP}_{n\mathbf{k}} = \epsilon_{n\mathbf{k}} + Z_{n\mathbf{k}} \big[
       \Sigma_{n\mathbf{k}}(\epsilon_{n\mathbf{k}})    - v^{xc}_{n\mathbf{k}} \big],
   \end{equation}
where $\epsilon_{n\mathbf{k}}$ are the KS eigenvalues, $\Sigma_{n\mathbf{k}}(\omega)$ and $v^{xc}_{n\mathbf{k}} $
are the expectation values of the self-energy and the exchange correlation potential, respectively, over the
$n\mathbf{k}$ KS-eigenvector, and $ Z_{n\mathbf{k}}$ the QP renormalization factor defined as:
 \begin{equation}
     Z_{n\mathbf{k}}  = \left[ 1 - \frac{\partial\Sigma_{n\mathbf{k}}
     (\omega)}{\partial\omega}\rvert_{\omega=\epsilon_{n\mathbf{k}}} \right]^{-1} .
 \end{equation}

The optical properties were calculated by solving the BSE whose kernel includes local field effects as well as the screened electron-hole interaction (with static screening). 
We also applied the Tamm-Dancoff approximation, which neglects the
coupling terms between resonant and anti-resonant blocks, after having numerically verified its validity. 

As an analysis tool, we computed atomically-projected DOS (pDOS) with the inclusion of GW QP corrections obtained
by Yambo, according to:
\begin{equation}
    \rho(\omega) = \frac{1}{N_{\mathbf{k}}} \sum_{\alpha} \sum_{n\mathbf{k}} \left| \langle \phi_\alpha |
    \psi_{n\mathbf{k}} \rangle  \right|^{2} \delta(\omega - E^{QP}_{n\mathbf{k}}),
    \label{eqn:PDOSQP}
\end{equation}
where $|\phi_\alpha\rangle$ are atomic orbitals.

Quasi particle unfolded band structures for the doped systems~\cite{pope-zung12prb,medeiros_unfolding_2015,medeiros_effects_2014} were calculated considering 
%taken from 
the spectral function $A(\boldsymbol{\kappa},\omega)$, resolved with respect to the $\boldsymbol{\kappa}$-vectors of the host pristine primitive cell (PC),
\begin{eqnarray}
A(\boldsymbol{\kappa},\omega) &=& \sum_{mn} \left| \langle \psi^{PC}_{m\boldsymbol{\kappa}} | \psi^{SC}_{n\mathbf{k}} \rangle \right|^2
\,\delta(\omega -E^{QP}_{n\mathbf{k}}),
\label{eq:unfolded_DOS}
\end{eqnarray}
where $\mathbf{k}=\mathbf{k}(\boldsymbol{\kappa})$ according to the folding induced by the supercell, and $ |\psi^{PC}_{m\boldsymbol{\kappa}}\rangle$'s form a complete set with $\boldsymbol{\kappa}$ symmetry.
Such band structures were obtained using the implementation of the unfolding procedure described in Refs.~[\onlinecite{medeiros_unfolding_2015,medeiros_effects_2014}].

%%%%%%%%%%%%%%%%%%%%%
\section{Results}
%%%%%%%%%%%%%%%%%%%%%
\label{sec:results}

\subsection{Structural properties}

\begin{figure}

    \centering
    \includegraphics[width=0.95\columnwidth]{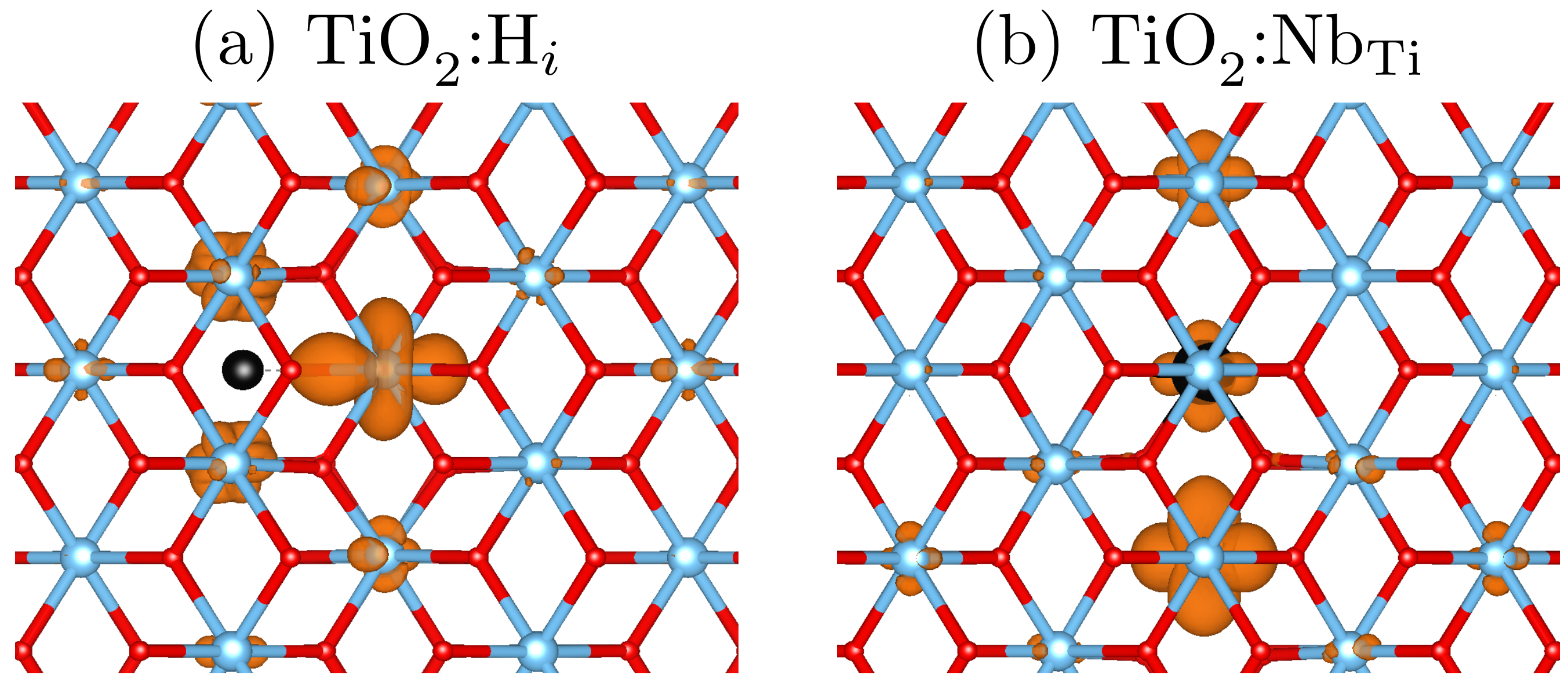}
    \caption{Relaxed structures of H (left) and Nb-doped (right) rutile TiO$_2$, viewed along the [010] direction. 
    The spin density isosurface (in orange) 
    is localized on the Ti  (and Nb) atoms near the defect center. Dopants (H and Nb) are  in black.
    }
    \label{fig:supercell:structure}
\end{figure}

The calculated structural parameters for pristine and doped rutile are summarized in Tab.~\ref{structural-details}.
In pristine rutile two slightly different  Ti-O bond lengths are present,  each
TiO$_6$ unit having two longer apical (type II, 1.99 {\AA}) and four shorter 
equatorial (type I, 1.96 {\AA}) Ti-O bonds.
In H-doped rutile (TiO$_2$:H$_{i}$), 
a large distortion of the TiO$_2$ crystal structure near the interstitial impurity 
takes place, with  type I bond lengths ranging from 1.90 up to 2.07 {\AA}, and   similar but less pronounced distortions of type II bonds. 
The H atom is bound to an oxygen with a bond length of 1.23 {\AA}  and is at distance 2.21 {\AA} from the closest Ti atom 
(Fig.~\ref{fig:supercell:structure}).
Significant distortions of the Ti-O bond lengths are found also for TiO$_2$:Nb$_{\textrm{Ti}}$, notably an elongation of the type I Ti-O bond for the first Ti atom adjacent to the defect (2.07 {\AA}) and a shortening for the subsequent bond  (1.90 {\AA}). Distortions of type II bonds are less significant in the case of substitutional Nb. As for Nb-O bonds, those of type I  are slightly shorter (1.94 {\AA}) than type I Ti-O bonds,  while those of type II are slightly longer  (2.02 {\AA}) than type II  Ti-O bonds. The distances between  Nb  and the closest Ti atoms are 2.94 and 3.04 {\AA}, to be compared to the Ti-Ti distance  of 2.97 {\AA} in  pristine rutile.

%%%%%%%%%%%%%%%%%%%%%%%%%%%%%%%%%%%%%%
\begin{table}
\centering
\caption{Bond lengths (in {\AA}) for pristine and doped rutile TiO$_2$. For the doped systems, maximum and minimum values of type I and type II  Ti-O bond lengths are given. Distances between the dopant and Ti (D-Ti) and O (D-O) are also reported.  
\label{structural-details}}
%
%\begin{ruledtabular}
\begin{tabular}{c|c|c|c}
{\bf bond}        &
{\bf TiO$_2$}     &
{\bf TiO$_2$:H$_{i}$}  &
{\bf TiO$_2$:Nb$_{\textrm{Ti}}$} \\[4pt]
\hline
%\hline
%
& & & \\[-5pt]
Ti-O    
& type I 1.96    &    1.91 - 2.07   &  1.90 - 2.07 \\%[5pt]
& type II 1.99   &    1.94 - 2.03   &  1.94 - 2.01  \\
%
%\hline
& & & \\[-5pt]
D-Ti    &      &   2.21  &  2.94 - 3.04     \\
%        &      &   2.21  &  \\%[5pt]
%
%\hline
& & & \\[-5pt]
D-O     &         &  1.23  &   1.93 - 2.01 \\
%        &         &  2.21  &    \\%[5pt]
%
\end{tabular}
%\end{ruledtabular}
%}
\end{table}
%%%%%%%%%%%%%%%%%%%%%%%%%%%%%%%%%%%%%%%%%%%%%%

\subsection{Band gap of pristine rutile}

For pristine rutile, our calculations give
a fundamental gap of 1.89 eV and 3.1 eV at PBE  and G$_0$W$_0$ levels, respectively. The latter value is in excellent agreement with calculations by Malashevich et al.\cite{PhysRevB.89.075205}(3.13 eV), who used a
G$_0$W$_0$[PBE] implementation 
within the complex generalized plasmon-pole approximation described in Ref.~\onlinecite{PhysRevB.40.3162}.

Similar values were obtained also by other G$_0$W$_0$ studies.
In particular,
Kang and Hybertsen~\cite{kang_quasiparticle_2010} obtained a band gap of 3.37 eV using a full-frequency contour-deformation (FF-CD) G$_0$W$_0$[PBE] approach; similarly, Zhang et
al.~\cite{zhang_all-electron_2016} found 3.30 eV by full-frequency G$_0$W$_0$[LDA]; Patrick and
Giustino\cite{patrick_gw_2012} obtained 3.40 eV at the G$_0$W$_0$[PBE] level with the GN-PPM;
Baldini et al\cite{baldini_anomalous_2017} obtained 3.30 eV using G$_0$W$_0$[PBE] with the PPM of Hybersten and Louie~\cite{PhysRevB.34.5390}; Chiodo et al,~\cite{chiodo_self-energy_2010} reported a gap of 3.59 eV using  G$_0$W$_0$[PBE] with GN-PPM.

Overall, the differences between our results and previous studies can be attributed to the different approach adopted to evaluate the frequency dependence of the self energy (Plasmon-pole Model PPM vs full-frequency methods and different flavours of PP model). In the case of Ref.~[\onlinecite{chiodo_self-energy_2010}], however, we explicitly verified that
differences from our results can be due to the pseudopotential dataset used in that work. 
In this regard, we notice that a recent study \cite{rang+18tobe} on the verification and validation of GW data
across three widely used GW codes (Yambo,\cite{mari+09cpc,sang+18tobe}
Abinit,\cite{Gonze2009}
and BerkeleyGW\cite{Deslippe2012}) found excellent agreement among the results obtained with these codes when
the same pseudopotential dataset was used. GN-PPM and FF-CD GW methods were also compared and found to lead to a
difference of the order of 0.1 eV for the fundamental gap of TiO$_2$ (changing from 3.2 eV with GN-PPM to 3.3 eV for FF-CD) when using (Fritz-Haber-Institute) FHI pseudopotentials with 12 valence electrons.

%========================
\subsection{Quasi-particle levels}
%======================== 

Figure~\ref{fig:supercell:structure} shows the unit cells of the relaxed structures of H- and Nb-doped TiO$_2$ together with the spin density associated with the donor states (in orange).   
We can see that for both H-  and Nb-doped rutile the donated electron  localizes  on the neighbouring Ti atoms, even though with different distributions. 
The  structural distortion close to the defect coupled to this density localization indicates the formation of localized polaronic states. 
Our results for H-doped rutile  agree with recent FTIR experiments~\cite{hupfer2017photoinduced} , which have provided evidence of  small electron polarons, each bound to a Ti atom adjacent to the OH group of the hydrogen defect. 

\begin{figure}
    \centering
    \includegraphics[width=0.95\columnwidth]{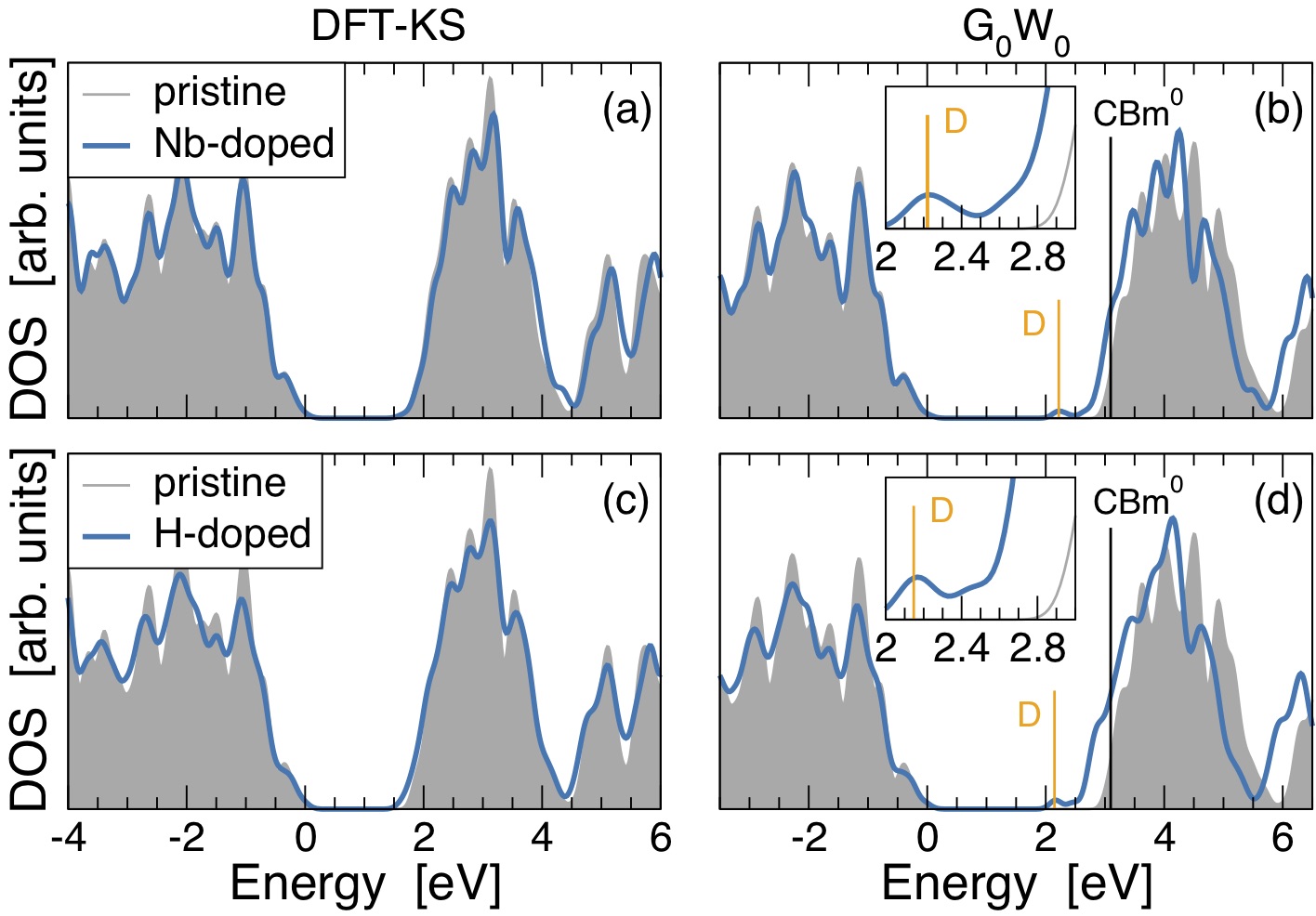}
    \caption KS-DFT (PBE) and G$_0$W$_0$ DOS for the pristine (grey  area), H-doped, and Nb-doped (blue curves) rutile TiO$_2$.  A Gaussian broadening of 0.01 Ry is used. In each panel, the energy is referred to the valence band maximum. Vertical lines indicate the position of the defect states (marked as D) and the CBm of pristine TiO$_2$ (marked as CBm$^0$). Spin resolved  G$_0$W$_0$ DOS are reported in the Supplementary Material Fig.~S7.
    \label{fig:rutile:pdos:ks-gw:doped}
\end{figure}

To evaluate the energy levels of the defect states in the TiO$_2$ band gap, we first examine the  Density of States (DOS), and align the different DOS at the 
valence band maximum (VBM), which is little affected by the dopants
%Projections of the DOS (pDOS) on TiO$_2$ and defect atoms are shown in 
(Fig.~\ref{fig:rutile:pdos:ks-gw:doped}).
At the KS-PBE level (left panels), defect states are found at the bottom of the TiO$_2$ conduction band. In the right panels, we show the same DOS after inclusion of quasi-particle corrections at the G$_0$W$_0$[PBE] level. 
In this case we clearly see 
defect states below the conduction bands, 
at 2.15 and 2.18 eV above the VBM for H and Nb, respectively. 
These energies correspond to -0.95 and -0.92 eV when referenced to the conduction band minimum 
(CBm) of pristine rutile.
Similar values have been reported in previous theoretical studies. For H-doped rutile, 
defect levels at
1.0~\cite{filippone_properties_2009} and
0.97 eV~\cite{stausholm-moller_dft+u_2010} 
below the CBm have been obtained using DFT+U, while hybrid HSE06 calculations predicted vertical defect ionization energies of 0.82 and 0.86 eV  for H- and Nb-doped rutile, respectively.\cite{deak_polaronic_2011}

The positions of the defect levels relative to the CBm 
%calculated in the presence of the dopant 
of the doped system are more difficult to assess,
because the CBm itself is not clearly defined due to 
dopant-TiO$_2$ hybridization effects and
an overall downshift of the conduction band.
%(-0.15/-0.2 eV for H~/~Nb~, see discussion below and
%
Besides the appearance of the defect states 
below the conduction bands (indicated with D), 
Figs.~\ref{fig:rutile:pdos:ks-gw:doped}(b,d) indeed show also an overall down-shift  of the conduction  band of TiO$_2$:H$_{i}$  and TiO$_2$:Nb$_{\textrm{Ti}}$ with respect to its position in pristine rutile (shaded area). 
%Fig.~\ref{fig:rutile:pdos:ks-gw:doped}(b). 
%As a result, the quasi-particle gap changes from 3.1 eV for  pristine rutile to 2.45  and 2.71 eV for %TiO$_2$:H$_{i}$  and TiO$_2$:Nb$_{\textrm{Ti}}$, respectively. 
This effect, already observed in Ref.~\onlinecite{chen2018origin}, is captured only after GW corrections and can be ascribed to enhanced screening effects due to the presence of the defects.

\begin{figure*}
    \centering
    \includegraphics[width=0.95\textwidth]{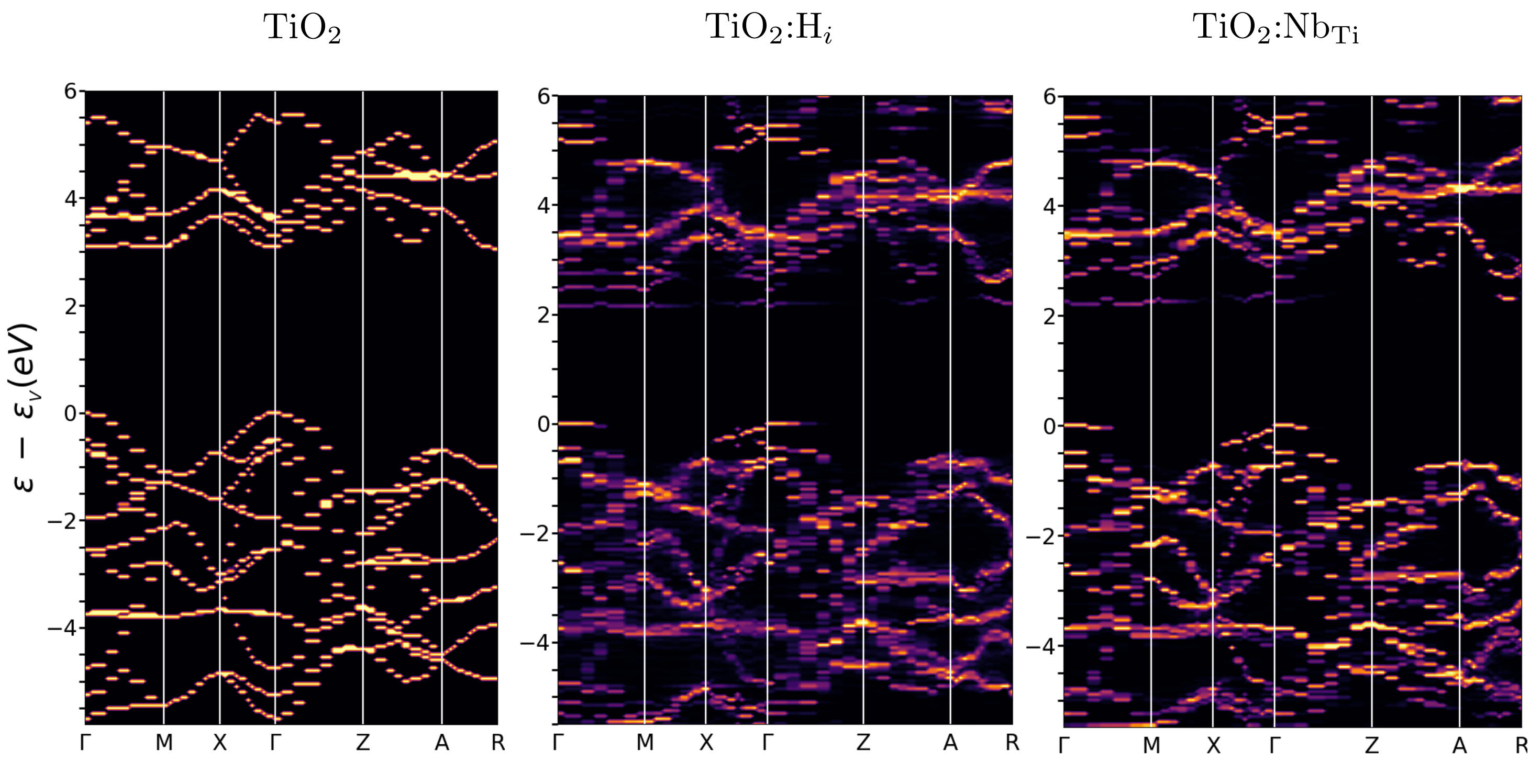}
    \caption{Unfolded quasi-particle band-structure %including G$_0$W$_0$ corrections 
    for pristine, 
    TiO$_2$:H$_{i}$  and TiO$_2$:Nb$_{\textrm{Ti}}$.
    Results for spin-up channels are shown; results for 
    spin down channels (which do not show any defect-related state) are reported in the Supplementary Material (Fig. S6).
    The color scale is such that black -- yellow correspond to [0,1] (in arbitrary units). }
    \label{fig:rutile:unfold:gw:doped}
\end{figure*}

In Fig.~\ref{fig:rutile:unfold:gw:doped} we show the unfolded quasi-particle band structures
for the doped systems, from Eq.~(\ref{eq:unfolded_DOS}),
compared to pristine rutile. We can recognize the presence of defect states below the CBm,  as indicated by the non-dispersive bands along selected symmetry lines 
that can be attributed to localized states.
Above these bands but still below the CBm of pristine TiO$_2$, additional blurry features are present.
The color scale represents the value of the  $\boldsymbol{\kappa}$-resolved DOS, as defined in Eq.~(\ref{eq:unfolded_DOS}): In each panel, black corresponds to zero and yellow to the maximum value of the DOS.  The sharper the DOS (i.e. the brighter the corresponding peak), the less hybridized the defect states. Within a mean field description, at fixed $\boldsymbol{\kappa}$-vector a delta peak would indeed show at the energy of each band if no defects were present. In the presence of defects instead, each peak is broadened because of the broken translational symmetry and the hybridization: the more blurred the picture, the larger the hybridization induced by the defect. 
The features above the non dispersive bands and below the CBm of pristine rutile can be thus attributed to such  strongly hybridized states.
Comparing  H- and Nb-doped TiO$_2$, a slightly higher degree of hybridization is apparent in TiO$_2$:H$_i$, which is consistent with the interstitial vs substitutional configuration of the defect. 

The quasi-particle band structures for the minority spin-down states of pristine and doped TiO$_2$ (Fig.~S6) and the spin resolved $G_0W_0$ density of states (Fig.~S7) are shown in the Supplemental Material. From Fig.~S6, it appears that the non-dispersive bands of defect states are now missing (as expected), and an overall   $\sim 0.2$ eV down-shift of the conduction bands of the doped systems relative to pristine rutile is present, due to the enhanced screening induced by the doping. We also notice that  the conduction band minima for the down-spin states of the doped systems are well-defined,  suggesting they may represent the  "true CBms" for the doped systems. This assumption,  supported also by Fig.~S7,  results in defect levels  lying at  -0.76 and -0.87 eV for  TiO$_2$:H$_{i}$  and TiO$_2$:Nb$_{\textrm{Ti}}$, respectively.

These results appear to disagree with the recent GW study by Chen et al.~\cite{chen2018origin}  which  concluded that the deep defect level at about 0.8-1 eV below the CBm of rutile can only originate from Ti interstitials, whereas oxygen vacancies and polarons can only give rise to shallow states.As pointed out also by Chen et al, the origin of the gap state has been  debated extensively in the experimental literature, where the current view is that both interstitials and oxygen vacancies contribute to the deep gap states.\cite{pang2013structure, yin2018excess} From a purely theoretical perspective, on the other hand, a firm conclusion may require  not only  well converged GW studies using different starting band-structures but also the study of different defect concentrations.

%========================
\subsection{Optical properties}
%========================

\begin{figure*}
    \centering
    \includegraphics[width=0.95\textwidth]{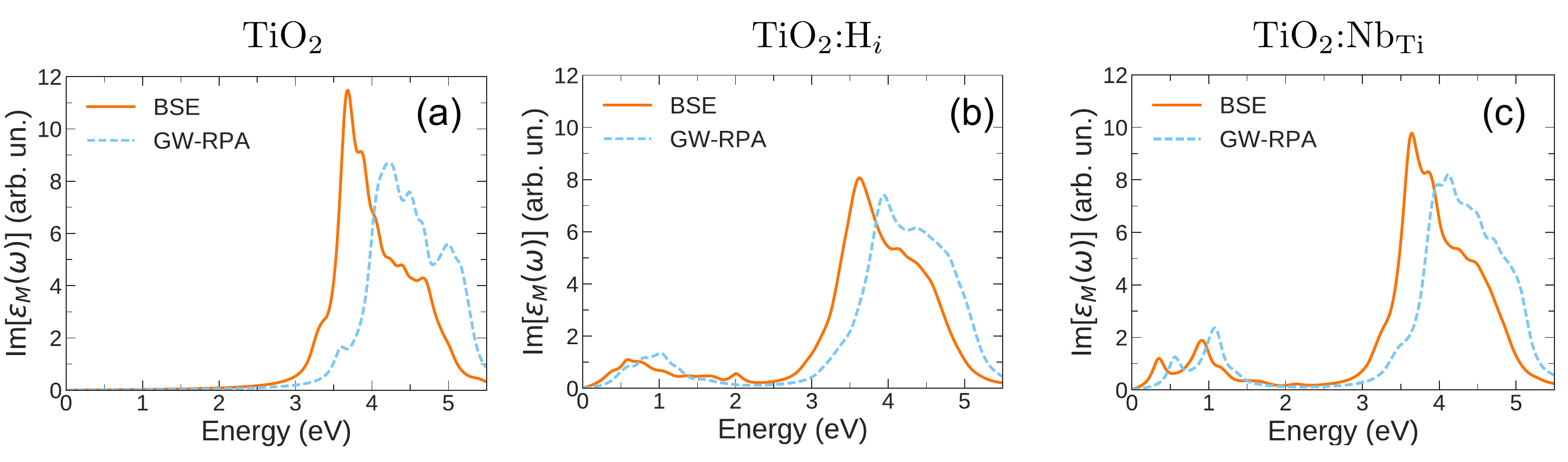}
   \caption{The imaginary part of the dielectric constant (averaged over x,y,z polarizations) for pristine, H-doped, and Nb-doped rutile, obtained by solving the  BSE and the independent quasi-particle spectra (GW-RPA).}
    \label{fig:rutile:bse:ip:opt} 
\end{figure*}

\begin{figure*}
    \centering
    \includegraphics[width=0.95\textwidth]{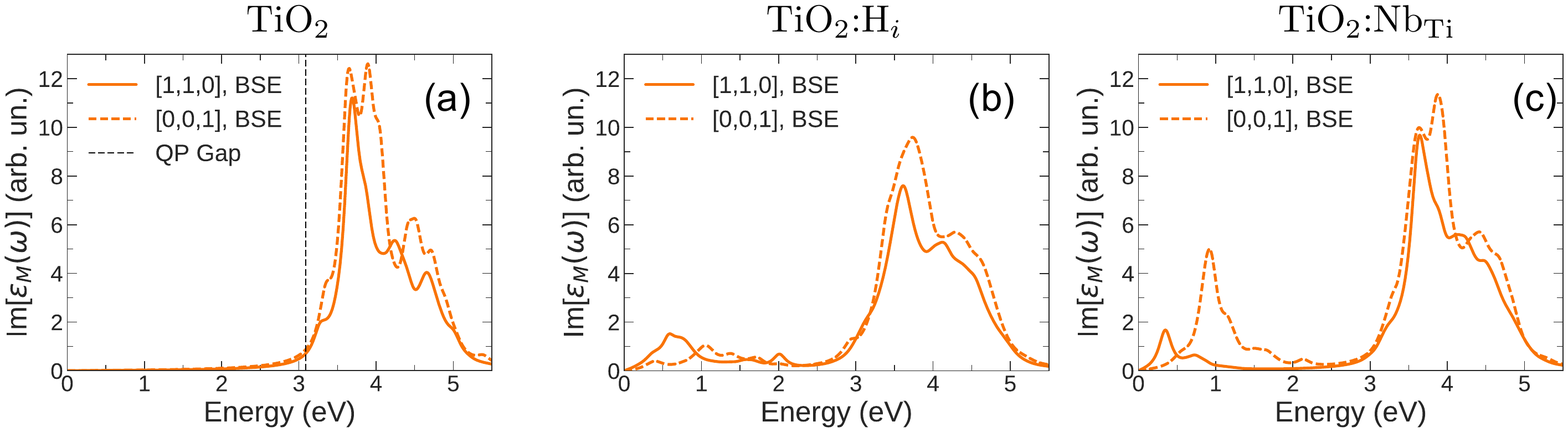}
    \caption{Optical anisotropy in pristine, H-doped, and Nb-doped rutile, obtained using BSE. E-field polarization with respect to the direct lattice vectors is indicated in the legend. 
    %Arrows indicate the excitons whose wavefunctions are shown in Fig. 6.
    \label{fig:rutile:anisotropy}}
\end{figure*}

%
{\it Pristine rutile.} Our computed BSE and independent particle (GW-RPA) spectra of pristine rutile  are  presented  in  Fig~\ref{fig:rutile:bse:ip:opt}(a).  The onset of the absorption in the BSE spectrum  is at $\sim 3.0$ eV, followed  by an intense narrow peak at 3.7 eV  and a shoulder at $\sim 4.4$ eV. 
Fig.~\ref{fig:rutile:anisotropy}(a)  further shows that the spectrum is anisotropic, with a characteristic double peaked structure for light polarized in the [001] direction, as already reported in previous calculations.~\cite{baldini_anomalous_2017,landmann_electronic_2012}
Both the onset and the shape of the calculated spectrum are in agreement with recent measurements,~\cite{baldini_anomalous_2017}
which show an onset at  $\sim 3$~eV, followed by a sharp peak at 3.93 eV, a shoulder at 4.51 eV and a broader peak at 5.42 eV. 
(We note that a larger number of  bands in BSE and more k-points
than  used in the present calculations 
would be  needed in order to reproduce also the high energy features of the experimental spectrum). 
Our results are also in line with recent~\cite{baldini_anomalous_2017,chiodo_self-energy_2010,landmann_electronic_2012} GW-BSE calculations: our
calculated red-shift of the first bright peak (0.53 eV) 
with respect to the independent quasi-particle picture  is comparable to the results of Chiodo {\it et al.} ~\cite{chiodo_self-energy_2010} and to the redshift in the computed spectra of Baldini {\it et al.}.~\cite{baldini_anomalous_2017} 
Differences observed for the excitation energies are due mainly to the different value of the computed fundamental gap (see discussion above). 

{\it Doped rutile.} 
Upon doping, new features in the absorption spectra appear, as shown in panels (b) and (c) of Figs.~\ref{fig:rutile:bse:ip:opt} and ~\ref{fig:rutile:anisotropy}. The most evident hallmark of the doped systems is the appearance of absorption peaks at low energy below 2 eV.

The low energy part of the spectrum of TiO$_2$:H$_{i}$ is characterized by a broad peak that  has a maximum at 0.6 eV and extends up to 2 eV. This peak  involves transitions from electrons in the gap states to the conduction bands. At higher energies, in the region of the spectrum where pristine rutile absorbs, we observe a small red-shift of the onset and the maximum of the absorption band with respect to the pristine crystal (from 3.68 to 3.62 eV for the maximum).
%AS I don't think the sentence below is correct
%as a consequence of the gap narrowing discussed above. 
%MJC I agree and have taken the sentence out
We also note that the shift of the BSE with respect to the independent particle spectrum is smaller in TiO$_2$:H$_{i}$ in comparison to that found for  pristine rutile, indicating an increased screening due to the excess electrons of the dopants. 
A red-shift of about 0.2 eV with respect to the independent particle spectrum is also observed in the lowest-energy, defect induced part of the spectrum, indicating the formation of slightly bound excitons.

%From the optical spectrum of H-doped rutile, Fig~\ref{fig:rutile:bse:ip:opt}(b), we observe an overall redshift of the main peak in comparison to the undoped material; in particular the onset shifts from 3.1 eV in pristine rutile to 2.8 eV in the doped case.  We also observe the appearance of new features in the spectra at about 1 eV, while the binding energy of the lowest exciton is reduced from ~0.5 to 0.3 eV. This can be attributed to an increased screening associated with the electrons of the dopants. 
%

In the case of Nb-doped rutile, Fig.~\ref{fig:rutile:bse:ip:opt}(c), the spectrum in the low energy region shows a defect induced absorption  characterized by a double peaked structure with maxima at 0.35 and 0.92 eV, sharper than those found for TiO$_2$:H$_{i}$.  
These excitations involve transitions from the defect state to conduction bands at different energies (see Figs. S3 and S4 of the Supp. Info.), and as in the case of TiO$_2$:H$_{i}$ the BSE spectrum exhibits a redshift with respect to the independent particle one, indicating the formation of bound excitons having a binding energy of 0.2 and 0.16 eV for the first and second peak respectively.
%double peakea new intense feature at 0.9 eV, much sharper than the diffuse features induced by Hydrogen doping. 
 As in the case of hydrogen doping, in the higher energy region we observe again a small red-shift of the onset and the maximum of the absorption (3.64 eV) with respect to pristine rutile. More importantly, also for TiO$_2$:Nb$_{\textrm{Ti}}$ the shift of the BSE spectrum with respect to the independent particle calculation (0.3 eV) is reduced in comparison to the pristine case.  
%While for TiO$_2$:Nb$_{\textrm{Ti}}$ we do not observe a significant redshift of the onset of the main peak at 3.1 eV, as found for H-doped rutile, we still find a reduction of the exciton binding energy to 0.3 eV.
Apart from the small rigid shift and an overall broadening, the shape of  the absorption spectrum of rutile in the high energy region is not significantly affected by the dopants.
%
%We  further notice a significant change in the general shape of the spectrum beyond the maximum intensity peak for both H- and Nb-doped rutile in comparison to pristine TiO$_2$. Moreover,  the main peak around 3.6 eV is  weaker for both dopants, with an overall broadening that suggests states of resonant character.
%

Figure~\ref{fig:rutile:anisotropy} shows the absorption spectra of  pristine and doped rutile for light polarized along and orthogonal to the {\it c} directions. 
Pristine rutile exhibits a significant optical anisotropy, associated with its tetragonal symmetry, which has been extensively studied both theoretically and experimentally.~\cite{baldini_anomalous_2017}
The anisotropy  is present also for the doped systems, as expected,  and is more pronounced in the low energy part of the spectra. In particular,   the two low energy peaks  in TiO$_2$:Nb$_{\textrm{Ti}}$ correspond to excitations that are optically active for light polarized along orthogonal directions. 
In order to verify that this anisotropy is not an artifact 
related to the small size of our supercell,
we performed additional DFT-KS calculations  using a  $(3\times 3 \times 4)$  supercell with 216 atoms. The results confirmed the presence of a strong anisotropy in the low energy part of the independent particle spectrum (see Fig. S8 in the SI).

\begin{figure}
    \centering
\includegraphics[width=.95\columnwidth]{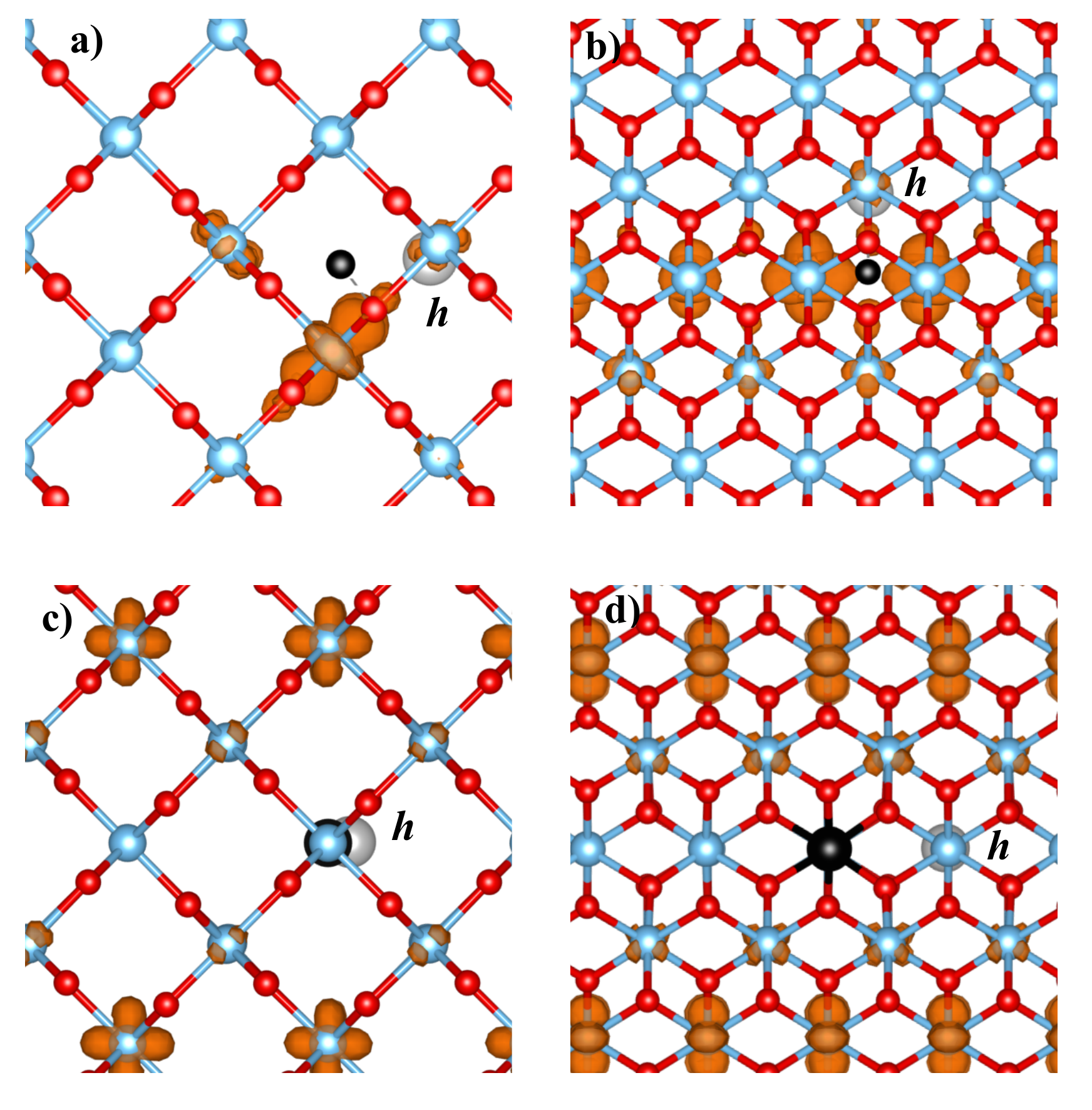}
    \caption{Excitonic wavefunctions corresponding to the absorption peaks at 0.33  eV in TiO$_2$:H$_{i}$  (a,b) and  0.35 eV in  TiO$_2$:Nb$_{\textrm{Ti}}$ (c,d). The view is along  [001] in  (a,c),  and along [010] in (b,d).  H and Niobium are shown in black; The position of the hole is indicated by a grey sphere and marked with \it{h}.
     %the gray ball indicates the position of the hole in the BSE calculations .
     %The view is along the  [001] direction
     }
    \label{fig:exx:wafefuctions}
\end{figure}

Figure~\ref{fig:exx:wafefuctions} shows the real space excitonic wave functions corresponding to the lowest energy  excitations of TiO$_2$:H$_{i}$ (top panel) and TiO$_2$:Nb$_{\textrm{Ti}}$ (bottom panel), which have very similar energies, 0.33  eV in TiO$_2$:H$_{i}$ and  0.35 eV in TiO$_2$:Nb$_{\textrm{Ti}}$. 
In order to obtain the exciton density, in both panels of Figure~\ref{fig:exx:wafefuctions} the position of the hole (indicated by {\it h}) was chosen in a region of high density of the occupied states mainly contributing to the excitation (see Figs.~S1 and S3 in Supp. Info).
We see that these excitations have very different spatial distributions in the two systems: while in TiO$_2$:H$_{i}$ the electron is strongly localized on the Ti atom closest to the defect, it is  almost completely delocalized in TiO$_2$:Nb$_{\textrm{Ti}}$. In TiO$_2$:H$_{i}$ the exciton is indeed composed of transitions from the gap state to conduction states mainly localized on the adjacent Ti atoms (see Fig. S1), whereas in TiO$_2$:Nb$_{\textrm{Ti}}$ it originates completely from the transition from the defect state to the lowest conduction band that is delocalized over Ti rows away from the defect (see Fig. S3). 
These characteristics are not restricted to the lowest excitations. A similar behaviour is indeed found also for other excitons at low energies: a localized wavefunction at 0.57 eV in TiO$_2$:H$_{i}$ and a delocalized exciton at 0.94 eV in TiO$_2$:Nb$_{\textrm{Ti}}$ (see Figs. S2 and S4 in SI).  
At higher energy, the excitonic wavefunctions  corresponding to the bright peak at 3.62 and 3.64 eV in TiO$_2$:H$_{i}$ and TiO$_2$:Nb$_{\textrm{Ti}}$ (Fig.S5) are delocalized, with a shape very similar to that of the exciton at 3.68 eV in pristine rutile, confirming the small impact of dopants on the absorption spectra in the high energy region.  

%%%%%%%%%%%%%%%%%%%%%
\section{Conclusions}
\label{sec:conclusions}
%%%%%%%%%%%%%%%%%%%%%
The introduction of dopants into the rutile TiO$_2$ lattice leads to significant modification of the
materials electronic and optical properties. 
Both interstitial Hydrogen and substitutional Niobium create a distortion of the crystal lattice around the defect and introduce electronic states localized mainly on adjacent Ti atoms. 
The resulting quasi-particle defect states form non dispersive bands with energies in the band gap  at 
$\sim 2.2$  eV above the VBM, which corresponds to $\sim -0.9$ eV relative to the CBm of pristine rutile.  The positions of the defect states relative to the CBm of  doped rutile are more difficult to estimate  due to  
the hybridization between dopant and Ti states, combined with an overall down-shift of the conduction band.  
 A reasonable choice is however to take  the CBm for the minority spin channel, which is well-defined in our calculations,  as the "true CBm"  for the doped system.  With this reference,  our results predict  the defect states  to be at $\sim$ -0.8 eV below the CBm,  consistent with experimental observations.
%The use of G$_0$W$_0$ is crucial for the accurate determination of the defect level positions, that are poorly described by DFT-KS, as well as for the description of the band gap narrowing upon doping. 

While the high energy part of the optical spectrum ($>3$ eV) is little affected by the presence of the dopant states, transition from the gap states to the conduction band give rise to new absorption peaks at low energy ($<2$ eV) that
enhance the absorption of rutile in the solar spectrum range. 
Interestingly, we found very different  excitonic wavefunctions for the low energy excitations  in H- and Nb doped rutile:  while photoexcited electrons are largely delocalized in  TiO$_2$:Nb$_{\textrm{Ti}}$,  they remain localized close to the hole in TiO$_2$:H$_{i}$. These characteristics suggest  a longer exciton lifetime and a larger mobility of photoexcited electrons in TiO$_2$:Nb$_{\textrm{Ti}}$, consistent with the use of Nb as an efficient dopant for improving the performance of TiO$_{2}$ in technological applications.\\
%%%%%%%%%%%%%%%%%%%%%
\section{Acknowledgments}
%%%%%%%%%%%%%%%%%%%%%
We thank Ivan Marri for helpful discussions about pseudopotential accuracy in describing TiO$_2$. 
We acknowledge support from the European Union H2020-INFRAEDI-2018-1 programme under grant agreement No. 824143 project "MaX - materials design at the exascale". 
%from the EU Centre of Excellence ``MaX - Materials Design at the Exascale'' (Horizon 2020 EINFRA-5, Grant No. 676598).
%we benefit in particular from the advancements in the Quantum ESPRESSO\cite{gian+09jpcm,gian+17jpcm} and Yambo\cite{mari+09cpc,sang+18tobe} codes, 
We acknowledge the use of the AiiDA\cite{PIZZI2016218} platform, a tool for automated high throughput computational science.
Computational resources were partly provided by PRACE (Grant No. Pra12\_3100) on the Fermi and Marconi machines at CINECA. M.J.C. acknowledges support from Brazilian Ministry of Science and Technology grant INEO, and  Brazilian National Council  for Scientific and Technological Development (CnPq) Brazil,  and CNR-Istituto Nanoscienze, Modena, Italy.
%
%\appendix
%\input{appendix}

%
% biblio
%
%\bibliographystyle{achemso}
%\bibliographystyle{aip}
%\bibliography{refs}

\begin{thebibliography}{61}
\expandafter\ifx\csname natexlab\endcsname\relax\def\natexlab#1{#1}\fi
\expandafter\ifx\csname bibnamefont\endcsname\relax
  \def\bibnamefont#1{#1}\fi
\expandafter\ifx\csname bibfnamefont\endcsname\relax
  \def\bibfnamefont#1{#1}\fi
\expandafter\ifx\csname citenamefont\endcsname\relax
  \def\citenamefont#1{#1}\fi
\expandafter\ifx\csname url\endcsname\relax
  \def\url#1{\texttt{#1}}\fi
\expandafter\ifx\csname urlprefix\endcsname\relax\def\urlprefix{URL }\fi
\providecommand{\bibinfo}[2]{#2}
\providecommand{\eprint}[2][]{\url{#2}}

\bibitem[{\citenamefont{Gr{\"a}tzel}(2001)}]{gratzel2001photoelectrochemical}
\bibinfo{author}{\bibfnamefont{M.}~\bibnamefont{Gr{\"a}tzel}},
  \bibinfo{journal}{nature} \textbf{\bibinfo{volume}{414}},
  \bibinfo{pages}{338} (\bibinfo{year}{2001}).

\bibitem[{\citenamefont{Diebold}(2003)}]{diebold2003surface}
\bibinfo{author}{\bibfnamefont{U.}~\bibnamefont{Diebold}},
  \bibinfo{journal}{Surface science reports} \textbf{\bibinfo{volume}{48}},
  \bibinfo{pages}{53} (\bibinfo{year}{2003}).

\bibitem[{\citenamefont{Asahi et~al.}(2001)\citenamefont{Asahi, Morikawa,
  Ohwaki, Aoki, and Taga}}]{asahi_visible-light_2001}
\bibinfo{author}{\bibfnamefont{R.}~\bibnamefont{Asahi}},
  \bibinfo{author}{\bibfnamefont{T.}~\bibnamefont{Morikawa}},
  \bibinfo{author}{\bibfnamefont{T.}~\bibnamefont{Ohwaki}},
  \bibinfo{author}{\bibfnamefont{K.}~\bibnamefont{Aoki}}, \bibnamefont{and}
  \bibinfo{author}{\bibfnamefont{Y.}~\bibnamefont{Taga}},
  \bibinfo{journal}{Science} \textbf{\bibinfo{volume}{293}},
  \bibinfo{pages}{269} (\bibinfo{year}{2001}).

\bibitem[{\citenamefont{Di~Valentin et~al.}(2005)\citenamefont{Di~Valentin,
  Pacchioni, Selloni, Livraghi, and
  Giamello}}]{di_valentin_characterization_2005}
\bibinfo{author}{\bibfnamefont{C.}~\bibnamefont{Di~Valentin}},
  \bibinfo{author}{\bibfnamefont{G.}~\bibnamefont{Pacchioni}},
  \bibinfo{author}{\bibfnamefont{A.}~\bibnamefont{Selloni}},
  \bibinfo{author}{\bibfnamefont{S.}~\bibnamefont{Livraghi}}, \bibnamefont{and}
  \bibinfo{author}{\bibfnamefont{E.}~\bibnamefont{Giamello}},
  \bibinfo{journal}{J. Phys. Chem. B} \textbf{\bibinfo{volume}{109}},
  \bibinfo{pages}{11414} (\bibinfo{year}{2005}).

\bibitem[{\citenamefont{Di~Valentin et~al.}(2009)\citenamefont{Di~Valentin,
  Pacchioni, and Selloni}}]{di_valentin_reduced_2009}
\bibinfo{author}{\bibfnamefont{C.}~\bibnamefont{Di~Valentin}},
  \bibinfo{author}{\bibfnamefont{G.}~\bibnamefont{Pacchioni}},
  \bibnamefont{and} \bibinfo{author}{\bibfnamefont{A.}~\bibnamefont{Selloni}},
  \bibinfo{journal}{J. Phys. Chem. C} \textbf{\bibinfo{volume}{113}},
  \bibinfo{pages}{20543} (\bibinfo{year}{2009}).

\bibitem[{\citenamefont{Di~Valentin and Pacchioni}(2013)}]{di2013trends}
\bibinfo{author}{\bibfnamefont{C.}~\bibnamefont{Di~Valentin}} \bibnamefont{and}
  \bibinfo{author}{\bibfnamefont{G.}~\bibnamefont{Pacchioni}},
  \bibinfo{journal}{Catalysis today} \textbf{\bibinfo{volume}{206}},
  \bibinfo{pages}{12} (\bibinfo{year}{2013}).

\bibitem[{\citenamefont{Sahasrabudhe et~al.}(2016)\citenamefont{Sahasrabudhe,
  Krizan, Bergman, Cava, and Schwartz}}]{doi:10.1021/acs.chemmater.6b02031}
\bibinfo{author}{\bibfnamefont{G.}~\bibnamefont{Sahasrabudhe}},
  \bibinfo{author}{\bibfnamefont{J.}~\bibnamefont{Krizan}},
  \bibinfo{author}{\bibfnamefont{S.~L.} \bibnamefont{Bergman}},
  \bibinfo{author}{\bibfnamefont{R.~J.} \bibnamefont{Cava}}, \bibnamefont{and}
  \bibinfo{author}{\bibfnamefont{J.}~\bibnamefont{Schwartz}},
  \bibinfo{journal}{Chem. Mater.} \textbf{\bibinfo{volume}{28}},
  \bibinfo{pages}{3630} (\bibinfo{year}{2016}).

\bibitem[{\citenamefont{Morris et~al.}(2000)\citenamefont{Morris, Dou, Rebane,
  Mitchell, Egdell, Law, Vittadini, and Casarin}}]{morris2000photoemission}
\bibinfo{author}{\bibfnamefont{D.}~\bibnamefont{Morris}},
  \bibinfo{author}{\bibfnamefont{Y.}~\bibnamefont{Dou}},
  \bibinfo{author}{\bibfnamefont{J.}~\bibnamefont{Rebane}},
  \bibinfo{author}{\bibfnamefont{C.}~\bibnamefont{Mitchell}},
  \bibinfo{author}{\bibfnamefont{R.}~\bibnamefont{Egdell}},
  \bibinfo{author}{\bibfnamefont{D.}~\bibnamefont{Law}},
  \bibinfo{author}{\bibfnamefont{A.}~\bibnamefont{Vittadini}},
  \bibnamefont{and} \bibinfo{author}{\bibfnamefont{M.}~\bibnamefont{Casarin}},
  \bibinfo{journal}{Physical Review B} \textbf{\bibinfo{volume}{61}},
  \bibinfo{pages}{13445} (\bibinfo{year}{2000}).

\bibitem[{\citenamefont{Stavola et~al.}(2014)\citenamefont{Stavola, Bekisli,
  Yin, Smithe, Beall~Fowler, and Boatner}}]{stavola2014contrasting}
\bibinfo{author}{\bibfnamefont{M.}~\bibnamefont{Stavola}},
  \bibinfo{author}{\bibfnamefont{F.}~\bibnamefont{Bekisli}},
  \bibinfo{author}{\bibfnamefont{W.}~\bibnamefont{Yin}},
  \bibinfo{author}{\bibfnamefont{K.}~\bibnamefont{Smithe}},
  \bibinfo{author}{\bibfnamefont{W.}~\bibnamefont{Beall~Fowler}},
  \bibnamefont{and} \bibinfo{author}{\bibfnamefont{L.~A.}
  \bibnamefont{Boatner}}, \bibinfo{journal}{Journal of Applied Physics}
  \textbf{\bibinfo{volume}{115}}, \bibinfo{pages}{012001}
  (\bibinfo{year}{2014}).

\bibitem[{\citenamefont{Sezen et~al.}(2014)\citenamefont{Sezen, Buchholz,
  Nefedov, Natzeck, Heissler, Di~Valentin, and W{\"o}ll}}]{sezen2014probing}
\bibinfo{author}{\bibfnamefont{H.}~\bibnamefont{Sezen}},
  \bibinfo{author}{\bibfnamefont{M.}~\bibnamefont{Buchholz}},
  \bibinfo{author}{\bibfnamefont{A.}~\bibnamefont{Nefedov}},
  \bibinfo{author}{\bibfnamefont{C.}~\bibnamefont{Natzeck}},
  \bibinfo{author}{\bibfnamefont{S.}~\bibnamefont{Heissler}},
  \bibinfo{author}{\bibfnamefont{C.}~\bibnamefont{Di~Valentin}},
  \bibnamefont{and} \bibinfo{author}{\bibfnamefont{C.}~\bibnamefont{W{\"o}ll}},
  \bibinfo{journal}{Scientific reports} \textbf{\bibinfo{volume}{4}},
  \bibinfo{pages}{3808} (\bibinfo{year}{2014}).

\bibitem[{\citenamefont{Hupfer et~al.}(2017{\natexlab{a}})\citenamefont{Hupfer,
  Vines, Monakhov, Svensson, and Herklotz}}]{hupfer2017photoinduced}
\bibinfo{author}{\bibfnamefont{A.}~\bibnamefont{Hupfer}},
  \bibinfo{author}{\bibfnamefont{L.}~\bibnamefont{Vines}},
  \bibinfo{author}{\bibfnamefont{E.}~\bibnamefont{Monakhov}},
  \bibinfo{author}{\bibfnamefont{B.}~\bibnamefont{Svensson}}, \bibnamefont{and}
  \bibinfo{author}{\bibfnamefont{F.}~\bibnamefont{Herklotz}},
  \bibinfo{journal}{Physical Review B} \textbf{\bibinfo{volume}{96}},
  \bibinfo{pages}{085203} (\bibinfo{year}{2017}{\natexlab{a}}).

\bibitem[{\citenamefont{Gerosa et~al.}(2017)\citenamefont{Gerosa, Bottani,
  Di~Valentin, Onida, and Pacchioni}}]{gerosa2017accuracy}
\bibinfo{author}{\bibfnamefont{M.}~\bibnamefont{Gerosa}},
  \bibinfo{author}{\bibfnamefont{C.}~\bibnamefont{Bottani}},
  \bibinfo{author}{\bibfnamefont{C.}~\bibnamefont{Di~Valentin}},
  \bibinfo{author}{\bibfnamefont{G.}~\bibnamefont{Onida}}, \bibnamefont{and}
  \bibinfo{author}{\bibfnamefont{G.}~\bibnamefont{Pacchioni}},
  \bibinfo{journal}{Journal of Physics: Condensed Matter}
  \textbf{\bibinfo{volume}{30}}, \bibinfo{pages}{044003}
  (\bibinfo{year}{2017}).

\bibitem[{\citenamefont{Thulin and Guerra}(2008)}]{thulin_calculations_2008}
\bibinfo{author}{\bibfnamefont{L.}~\bibnamefont{Thulin}} \bibnamefont{and}
  \bibinfo{author}{\bibfnamefont{J.}~\bibnamefont{Guerra}},
  \bibinfo{journal}{Phys. Rev. B} \textbf{\bibinfo{volume}{77}},
  \bibinfo{pages}{195112} (\bibinfo{year}{2008}).

\bibitem[{\citenamefont{Chiodo et~al.}(2011)\citenamefont{Chiodo,
  Garc{\'\i}a-Lastra, Mowbray, Rubio, and Iacomino}}]{chiodo_tailoring_2010}
\bibinfo{author}{\bibfnamefont{L.}~\bibnamefont{Chiodo}},
  \bibinfo{author}{\bibfnamefont{J.~M.} \bibnamefont{Garc{\'\i}a-Lastra}},
  \bibinfo{author}{\bibfnamefont{D.~J.} \bibnamefont{Mowbray}},
  \bibinfo{author}{\bibfnamefont{A.}~\bibnamefont{Rubio}}, \bibnamefont{and}
  \bibinfo{author}{\bibfnamefont{A.}~\bibnamefont{Iacomino}}, in
  \emph{\bibinfo{booktitle}{Computational Studies of New Materials II: From
  Ultrafast Processes and Nanostructures to Optoelectronics, Energy Storage and
  Nanomedicine}} (\bibinfo{publisher}{World Scientific}, \bibinfo{year}{2011}),
  pp. \bibinfo{pages}{301--329}.

\bibitem[{\citenamefont{Kang and Hybertsen}(2010)}]{kang_quasiparticle_2010}
\bibinfo{author}{\bibfnamefont{W.}~\bibnamefont{Kang}} \bibnamefont{and}
  \bibinfo{author}{\bibfnamefont{M.~S.} \bibnamefont{Hybertsen}},
  \bibinfo{journal}{Phys Rev B} \textbf{\bibinfo{volume}{82}},
  \bibinfo{pages}{085203} (\bibinfo{year}{2010}).

\bibitem[{\citenamefont{{Alves-Santos}
  et~al.}(2014)\citenamefont{{Alves-Santos}, Jorge, Caldas, and
  Varsano}}]{alves-santos_electronic_2014}
\bibinfo{author}{\bibfnamefont{M.}~\bibnamefont{{Alves-Santos}}},
  \bibinfo{author}{\bibfnamefont{L.~M.~M.} \bibnamefont{Jorge}},
  \bibinfo{author}{\bibfnamefont{M.~J.} \bibnamefont{Caldas}},
  \bibnamefont{and} \bibinfo{author}{\bibfnamefont{D.}~\bibnamefont{Varsano}},
  \bibinfo{journal}{J. Phys. Chem. C} \textbf{\bibinfo{volume}{118}},
  \bibinfo{pages}{13539} (\bibinfo{year}{2014}).

\bibitem[{\citenamefont{Kohn and Sham}(1965)}]{kohn-sham65pr}
\bibinfo{author}{\bibfnamefont{W.}~\bibnamefont{Kohn}} \bibnamefont{and}
  \bibinfo{author}{\bibfnamefont{L.~J.} \bibnamefont{Sham}},
  \bibinfo{journal}{Phys. Rev.} \textbf{\bibinfo{volume}{140}},
  \bibinfo{pages}{A1133} (\bibinfo{year}{1965}).

\bibitem[{\citenamefont{Giannozzi et~al.}(2009)\citenamefont{Giannozzi, Baroni,
  Bonini, Calandra, Car, Cavazzoni, {Davide Ceresoli}, Chiarotti, Cococcioni,
  Dabo et~al.}}]{giannozzi_quantum_2009}
\bibinfo{author}{\bibfnamefont{P.}~\bibnamefont{Giannozzi}},
  \bibinfo{author}{\bibfnamefont{S.}~\bibnamefont{Baroni}},
  \bibinfo{author}{\bibfnamefont{N.}~\bibnamefont{Bonini}},
  \bibinfo{author}{\bibfnamefont{M.}~\bibnamefont{Calandra}},
  \bibinfo{author}{\bibfnamefont{R.}~\bibnamefont{Car}},
  \bibinfo{author}{\bibfnamefont{C.}~\bibnamefont{Cavazzoni}},
  \bibinfo{author}{\bibnamefont{{Davide Ceresoli}}},
  \bibinfo{author}{\bibfnamefont{G.~L.} \bibnamefont{Chiarotti}},
  \bibinfo{author}{\bibfnamefont{M.}~\bibnamefont{Cococcioni}},
  \bibinfo{author}{\bibfnamefont{I.}~\bibnamefont{Dabo}}, \bibnamefont{et~al.},
  \bibinfo{journal}{J. Phys. Condens. Matter} \textbf{\bibinfo{volume}{21}},
  \bibinfo{pages}{395502} (\bibinfo{year}{2009}).

\bibitem[{\citenamefont{Giannozzi et~al.}(2017)\citenamefont{Giannozzi,
  Andreussi, Brumme, Bunau, Nardelli, Calandra, Car, Cavazzoni, {D Ceresoli},
  Cococcioni et~al.}}]{giannozzi_advanced_2017}
\bibinfo{author}{\bibfnamefont{P.}~\bibnamefont{Giannozzi}},
  \bibinfo{author}{\bibfnamefont{O.}~\bibnamefont{Andreussi}},
  \bibinfo{author}{\bibfnamefont{T.}~\bibnamefont{Brumme}},
  \bibinfo{author}{\bibfnamefont{O.}~\bibnamefont{Bunau}},
  \bibinfo{author}{\bibfnamefont{M.~B.} \bibnamefont{Nardelli}},
  \bibinfo{author}{\bibfnamefont{M.}~\bibnamefont{Calandra}},
  \bibinfo{author}{\bibfnamefont{R.}~\bibnamefont{Car}},
  \bibinfo{author}{\bibfnamefont{C.}~\bibnamefont{Cavazzoni}},
  \bibinfo{author}{\bibnamefont{{D Ceresoli}}},
  \bibinfo{author}{\bibfnamefont{M.}~\bibnamefont{Cococcioni}},
  \bibnamefont{et~al.}, \bibinfo{journal}{J. Phys. Condens. Matter}
  \textbf{\bibinfo{volume}{29}}, \bibinfo{pages}{465901}
  (\bibinfo{year}{2017}).

\bibitem[{\citenamefont{Perdew et~al.}(1996)\citenamefont{Perdew, Burke, and
  Ernzerhof}}]{perd+96prl}
\bibinfo{author}{\bibfnamefont{J.~P.} \bibnamefont{Perdew}},
  \bibinfo{author}{\bibfnamefont{K.}~\bibnamefont{Burke}}, \bibnamefont{and}
  \bibinfo{author}{\bibfnamefont{M.}~\bibnamefont{Ernzerhof}},
  \bibinfo{journal}{Phys. Rev. Lett.} \textbf{\bibinfo{volume}{77}},
  \bibinfo{pages}{3865} (\bibinfo{year}{1996}).

\bibitem[{\citenamefont{Hamann}(2013)}]{hamann_optimized_2013}
\bibinfo{author}{\bibfnamefont{D.~R.} \bibnamefont{Hamann}},
  \bibinfo{journal}{Phys Rev B} \textbf{\bibinfo{volume}{88}},
  \bibinfo{pages}{085117} (\bibinfo{year}{2013}).

\bibitem[{\citenamefont{Cococcioni and {de Gironcoli}}(2005)}]{coco-degi05prb}
\bibinfo{author}{\bibfnamefont{M.}~\bibnamefont{Cococcioni}} \bibnamefont{and}
  \bibinfo{author}{\bibfnamefont{S.}~\bibnamefont{{de Gironcoli}}},
  \bibinfo{journal}{Phys. Rev. B} \textbf{\bibinfo{volume}{71}},
  \bibinfo{pages}{035105} (\bibinfo{year}{2005}).

\bibitem[{\citenamefont{Finazzi et~al.}(2008)\citenamefont{Finazzi,
  Di~Valentin, Pacchioni, and Selloni}}]{finazzi_excess_2008}
\bibinfo{author}{\bibfnamefont{E.}~\bibnamefont{Finazzi}},
  \bibinfo{author}{\bibfnamefont{C.}~\bibnamefont{Di~Valentin}},
  \bibinfo{author}{\bibfnamefont{G.}~\bibnamefont{Pacchioni}},
  \bibnamefont{and} \bibinfo{author}{\bibfnamefont{A.}~\bibnamefont{Selloni}},
  \bibinfo{journal}{J. Chem. Phys.} \textbf{\bibinfo{volume}{129}},
  \bibinfo{pages}{154113} (\bibinfo{year}{2008}).

\bibitem[{\citenamefont{Morgan and Watson}(2007)}]{morgan_dft+u_2007}
\bibinfo{author}{\bibfnamefont{B.~J.} \bibnamefont{Morgan}} \bibnamefont{and}
  \bibinfo{author}{\bibfnamefont{G.~W.} \bibnamefont{Watson}},
  \bibinfo{journal}{Surf. Sci.} \textbf{\bibinfo{volume}{601}},
  \bibinfo{pages}{5034} (\bibinfo{year}{2007}).

\bibitem[{\citenamefont{Mattioli et~al.}(2008)\citenamefont{Mattioli,
  Filippone, Alippi, and Amore~Bonapasta}}]{PhysRevB.78.241201}
\bibinfo{author}{\bibfnamefont{G.}~\bibnamefont{Mattioli}},
  \bibinfo{author}{\bibfnamefont{F.}~\bibnamefont{Filippone}},
  \bibinfo{author}{\bibfnamefont{P.}~\bibnamefont{Alippi}}, \bibnamefont{and}
  \bibinfo{author}{\bibfnamefont{A.}~\bibnamefont{Amore~Bonapasta}},
  \bibinfo{journal}{Phys Rev B} \textbf{\bibinfo{volume}{78}},
  \bibinfo{pages}{241201} (\bibinfo{year}{2008}).

\bibitem[{\citenamefont{Setvin et~al.}(2014)\citenamefont{Setvin, Franchini,
  Hao, Schmid, Janotti, Kaltak, Van~de Walle, Kresse, and
  Diebold}}]{setvin2014direct}
\bibinfo{author}{\bibfnamefont{M.}~\bibnamefont{Setvin}},
  \bibinfo{author}{\bibfnamefont{C.}~\bibnamefont{Franchini}},
  \bibinfo{author}{\bibfnamefont{X.}~\bibnamefont{Hao}},
  \bibinfo{author}{\bibfnamefont{M.}~\bibnamefont{Schmid}},
  \bibinfo{author}{\bibfnamefont{A.}~\bibnamefont{Janotti}},
  \bibinfo{author}{\bibfnamefont{M.}~\bibnamefont{Kaltak}},
  \bibinfo{author}{\bibfnamefont{C.~G.} \bibnamefont{Van~de Walle}},
  \bibinfo{author}{\bibfnamefont{G.}~\bibnamefont{Kresse}}, \bibnamefont{and}
  \bibinfo{author}{\bibfnamefont{U.}~\bibnamefont{Diebold}},
  \bibinfo{journal}{Physical review letters} \textbf{\bibinfo{volume}{113}},
  \bibinfo{pages}{086402} (\bibinfo{year}{2014}).

\bibitem[{\citenamefont{Zhao et~al.}(2013)\citenamefont{Zhao, Hou, Li, Chan,
  and Lee}}]{zhao2013effective}
\bibinfo{author}{\bibfnamefont{Y.}~\bibnamefont{Zhao}},
  \bibinfo{author}{\bibfnamefont{T.}~\bibnamefont{Hou}},
  \bibinfo{author}{\bibfnamefont{Y.}~\bibnamefont{Li}},
  \bibinfo{author}{\bibfnamefont{K.~S.} \bibnamefont{Chan}}, \bibnamefont{and}
  \bibinfo{author}{\bibfnamefont{S.-T.} \bibnamefont{Lee}},
  \bibinfo{journal}{Appl. Phys. Lett.} \textbf{\bibinfo{volume}{102}},
  \bibinfo{pages}{171902} (\bibinfo{year}{2013}).

\bibitem[{\citenamefont{Dette et~al.}(2014)\citenamefont{Dette,
  {P\'erez-Osorio}, Kley, Punke, Patrick, Jacobson, Giustino, Jung, and
  Kern}}]{dette_tio2_2014}
\bibinfo{author}{\bibfnamefont{C.}~\bibnamefont{Dette}},
  \bibinfo{author}{\bibfnamefont{M.~A.} \bibnamefont{{P\'erez-Osorio}}},
  \bibinfo{author}{\bibfnamefont{C.~S.} \bibnamefont{Kley}},
  \bibinfo{author}{\bibfnamefont{P.}~\bibnamefont{Punke}},
  \bibinfo{author}{\bibfnamefont{C.~E.} \bibnamefont{Patrick}},
  \bibinfo{author}{\bibfnamefont{P.}~\bibnamefont{Jacobson}},
  \bibinfo{author}{\bibfnamefont{F.}~\bibnamefont{Giustino}},
  \bibinfo{author}{\bibfnamefont{S.~J.} \bibnamefont{Jung}}, \bibnamefont{and}
  \bibinfo{author}{\bibfnamefont{K.}~\bibnamefont{Kern}},
  \bibinfo{journal}{Nano Lett.} \textbf{\bibinfo{volume}{14}},
  \bibinfo{pages}{6533} (\bibinfo{year}{2014}).

\bibitem[{\citenamefont{Selcuk et~al.}(2018)\citenamefont{Selcuk, Zhao, and
  Selloni}}]{selcuk_structural_2018}
\bibinfo{author}{\bibfnamefont{S.}~\bibnamefont{Selcuk}},
  \bibinfo{author}{\bibfnamefont{X.}~\bibnamefont{Zhao}}, \bibnamefont{and}
  \bibinfo{author}{\bibfnamefont{A.}~\bibnamefont{Selloni}},
  \bibinfo{journal}{Nature Materials}  (\bibinfo{year}{2018}).

\bibitem[{\citenamefont{Aschauer and Selloni}(2012)}]{aschauer2012hydrogen}
\bibinfo{author}{\bibfnamefont{U.}~\bibnamefont{Aschauer}} \bibnamefont{and}
  \bibinfo{author}{\bibfnamefont{A.}~\bibnamefont{Selloni}},
  \bibinfo{journal}{Physical Chemistry Chemical Physics}
  \textbf{\bibinfo{volume}{14}}, \bibinfo{pages}{16595} (\bibinfo{year}{2012}).

\bibitem[{\citenamefont{Cromer and Herrington}(1955)}]{doi:10.1021/ja01623a004}
\bibinfo{author}{\bibfnamefont{D.~T.} \bibnamefont{Cromer}} \bibnamefont{and}
  \bibinfo{author}{\bibfnamefont{K.}~\bibnamefont{Herrington}},
  \bibinfo{journal}{J. Am. Chem. Soc.} \textbf{\bibinfo{volume}{77}},
  \bibinfo{pages}{4708} (\bibinfo{year}{1955}).

\bibitem[{\citenamefont{Onida et~al.}(2002)\citenamefont{Onida, Reining, and
  Rubio}}]{onid+02rmp}
\bibinfo{author}{\bibfnamefont{G.}~\bibnamefont{Onida}},
  \bibinfo{author}{\bibfnamefont{L.}~\bibnamefont{Reining}}, \bibnamefont{and}
  \bibinfo{author}{\bibfnamefont{A.}~\bibnamefont{Rubio}},
  \bibinfo{journal}{Rev. Mod. Phys.} \textbf{\bibinfo{volume}{74}},
  \bibinfo{pages}{601} (\bibinfo{year}{2002}).

\bibitem[{\citenamefont{Hanke and Sham}(1974)}]{Hanke74}
\bibinfo{author}{\bibfnamefont{W.}~\bibnamefont{Hanke}} \bibnamefont{and}
  \bibinfo{author}{\bibfnamefont{L.~J.} \bibnamefont{Sham}},
  \bibinfo{journal}{Phys Rev Lett} \textbf{\bibinfo{volume}{33}},
  \bibinfo{pages}{582} (\bibinfo{year}{1974}).

\bibitem[{\citenamefont{Hanke and Sham}(1980)}]{Hanke80}
\bibinfo{author}{\bibfnamefont{W.}~\bibnamefont{Hanke}} \bibnamefont{and}
  \bibinfo{author}{\bibfnamefont{L.~J.} \bibnamefont{Sham}},
  \bibinfo{journal}{Phys Rev B} \textbf{\bibinfo{volume}{21}},
  \bibinfo{pages}{4656} (\bibinfo{year}{1980}).

\bibitem[{\citenamefont{Strinati et~al.}(1982)\citenamefont{Strinati,
  Mattausch, and Hanke}}]{Strinati82}
\bibinfo{author}{\bibfnamefont{G.}~\bibnamefont{Strinati}},
  \bibinfo{author}{\bibfnamefont{H.~J.} \bibnamefont{Mattausch}},
  \bibnamefont{and} \bibinfo{author}{\bibfnamefont{W.}~\bibnamefont{Hanke}},
  \bibinfo{journal}{Phys Rev B(R)} \textbf{\bibinfo{volume}{25}},
  \bibinfo{pages}{2867} (\bibinfo{year}{1982}).

\bibitem[{\citenamefont{Strinati}(1988)}]{Strinati88}
\bibinfo{author}{\bibfnamefont{G.}~\bibnamefont{Strinati}},
  \bibinfo{journal}{Rivista Nuovo Cimento} \textbf{\bibinfo{volume}{11}},
  \bibinfo{pages}{1} (\bibinfo{year}{1988}).

\bibitem[{\citenamefont{Godby and Needs}(1989)}]{godb-need89prl}
\bibinfo{author}{\bibfnamefont{R.~W.} \bibnamefont{Godby}} \bibnamefont{and}
  \bibinfo{author}{\bibfnamefont{R.~J.} \bibnamefont{Needs}},
  \bibinfo{journal}{Phys. Rev. Lett.} \textbf{\bibinfo{volume}{62}},
  \bibinfo{pages}{1169} (\bibinfo{year}{1989}).

\bibitem[{\citenamefont{Marini et~al.}(2009)\citenamefont{Marini, Hogan,
  Gr{\"u}ning, and Varsano}}]{mari+09cpc}
\bibinfo{author}{\bibfnamefont{A.}~\bibnamefont{Marini}},
  \bibinfo{author}{\bibfnamefont{C.}~\bibnamefont{Hogan}},
  \bibinfo{author}{\bibfnamefont{M.}~\bibnamefont{Gr{\"u}ning}},
  \bibnamefont{and} \bibinfo{author}{\bibfnamefont{D.}~\bibnamefont{Varsano}},
  \bibinfo{journal}{Comput. Phys. Commun.} \textbf{\bibinfo{volume}{180}},
  \bibinfo{pages}{1392} (\bibinfo{year}{2009}).

\bibitem[{\citenamefont{Sangalli et~al.}(2018)\citenamefont{Sangalli, Hogan,
  Ferretti, Varsano, Gr\"uning, Palummo, Attaccalite, Cannuccia, Marsili,
  Affinito et~al.}}]{sang+18tobe}
\bibinfo{author}{\bibfnamefont{D.}~\bibnamefont{Sangalli}},
  \bibinfo{author}{\bibfnamefont{C.}~\bibnamefont{Hogan}},
  \bibinfo{author}{\bibfnamefont{A.}~\bibnamefont{Ferretti}},
  \bibinfo{author}{\bibfnamefont{D.}~\bibnamefont{Varsano}},
  \bibinfo{author}{\bibfnamefont{M.}~\bibnamefont{Gr\"uning}},
  \bibinfo{author}{\bibfnamefont{M.}~\bibnamefont{Palummo}},
  \bibinfo{author}{\bibfnamefont{C.}~\bibnamefont{Attaccalite}},
  \bibinfo{author}{\bibfnamefont{E.}~\bibnamefont{Cannuccia}},
  \bibinfo{author}{\bibfnamefont{M.}~\bibnamefont{Marsili}},
  \bibinfo{author}{\bibfnamefont{F.}~\bibnamefont{Affinito}},
  \bibnamefont{et~al.}, \bibinfo{journal}{preprint}  (\bibinfo{year}{2018}).

\bibitem[{\citenamefont{Popescu and Zunger}(2012)}]{pope-zung12prb}
\bibinfo{author}{\bibfnamefont{V.}~\bibnamefont{Popescu}} \bibnamefont{and}
  \bibinfo{author}{\bibfnamefont{A.}~\bibnamefont{Zunger}},
  \bibinfo{journal}{Phys. Rev. B} \textbf{\bibinfo{volume}{85}},
  \bibinfo{pages}{085201} (\bibinfo{year}{2012}).

\bibitem[{\citenamefont{Medeiros et~al.}(2015)\citenamefont{Medeiros, Tsirkin,
  Stafstr\"om, and Bj\"ork}}]{medeiros_unfolding_2015}
\bibinfo{author}{\bibfnamefont{P.~V.~C.} \bibnamefont{Medeiros}},
  \bibinfo{author}{\bibfnamefont{S.~S.} \bibnamefont{Tsirkin}},
  \bibinfo{author}{\bibfnamefont{S.}~\bibnamefont{Stafstr\"om}},
  \bibnamefont{and} \bibinfo{author}{\bibfnamefont{J.}~\bibnamefont{Bj\"ork}},
  \bibinfo{journal}{Phys Rev B} \textbf{\bibinfo{volume}{91}},
  \bibinfo{pages}{041116} (\bibinfo{year}{2015}).

\bibitem[{\citenamefont{Medeiros et~al.}(2014)\citenamefont{Medeiros,
  Stafstr\"om, and Bj\"ork}}]{medeiros_effects_2014}
\bibinfo{author}{\bibfnamefont{P.~V.~C.} \bibnamefont{Medeiros}},
  \bibinfo{author}{\bibfnamefont{S.}~\bibnamefont{Stafstr\"om}},
  \bibnamefont{and} \bibinfo{author}{\bibfnamefont{J.}~\bibnamefont{Bj\"ork}},
  \bibinfo{journal}{Phys Rev B} \textbf{\bibinfo{volume}{89}},
  \bibinfo{pages}{041407} (\bibinfo{year}{2014}).

\bibitem[{\citenamefont{Malashevich et~al.}(2014)\citenamefont{Malashevich,
  Jain, and Louie}}]{PhysRevB.89.075205}
\bibinfo{author}{\bibfnamefont{A.}~\bibnamefont{Malashevich}},
  \bibinfo{author}{\bibfnamefont{M.}~\bibnamefont{Jain}}, \bibnamefont{and}
  \bibinfo{author}{\bibfnamefont{S.~G.} \bibnamefont{Louie}},
  \bibinfo{journal}{Phys Rev B} \textbf{\bibinfo{volume}{89}},
  \bibinfo{pages}{075205} (\bibinfo{year}{2014}).

\bibitem[{\citenamefont{Zhang et~al.}(1989)\citenamefont{Zhang, Tom\'anek,
  Cohen, Louie, and Hybertsen}}]{PhysRevB.40.3162}
\bibinfo{author}{\bibfnamefont{S.~B.} \bibnamefont{Zhang}},
  \bibinfo{author}{\bibfnamefont{D.}~\bibnamefont{Tom\'anek}},
  \bibinfo{author}{\bibfnamefont{M.~L.} \bibnamefont{Cohen}},
  \bibinfo{author}{\bibfnamefont{S.~G.} \bibnamefont{Louie}}, \bibnamefont{and}
  \bibinfo{author}{\bibfnamefont{M.~S.} \bibnamefont{Hybertsen}},
  \bibinfo{journal}{Phys. Rev. B} \textbf{\bibinfo{volume}{40}},
  \bibinfo{pages}{3162} (\bibinfo{year}{1989}).

\bibitem[{\citenamefont{Zhang et~al.}(2016)\citenamefont{Zhang, Ono, Nagatsuka,
  and Ohno}}]{zhang_all-electron_2016}
\bibinfo{author}{\bibfnamefont{M.}~\bibnamefont{Zhang}},
  \bibinfo{author}{\bibfnamefont{S.}~\bibnamefont{Ono}},
  \bibinfo{author}{\bibfnamefont{N.}~\bibnamefont{Nagatsuka}},
  \bibnamefont{and} \bibinfo{author}{\bibfnamefont{K.}~\bibnamefont{Ohno}},
  \bibinfo{journal}{Phys. Rev. B} \textbf{\bibinfo{volume}{93}},
  \bibinfo{pages}{155116} (\bibinfo{year}{2016}).

\bibitem[{\citenamefont{Patrick and Giustino}(2012)}]{patrick_gw_2012}
\bibinfo{author}{\bibfnamefont{C.~E.} \bibnamefont{Patrick}} \bibnamefont{and}
  \bibinfo{author}{\bibfnamefont{F.}~\bibnamefont{Giustino}},
  \bibinfo{journal}{J. Phys. Condens. Matter} \textbf{\bibinfo{volume}{24}},
  \bibinfo{pages}{202201} (\bibinfo{year}{2012}).

\bibitem[{\citenamefont{Baldini et~al.}(2017)\citenamefont{Baldini, Dominguez,
  Chiodo, Sheveleva, {Yazdi-Rizi}, Bernhard, Rubio, and
  Chergui}}]{baldini_anomalous_2017}
\bibinfo{author}{\bibfnamefont{E.}~\bibnamefont{Baldini}},
  \bibinfo{author}{\bibfnamefont{A.}~\bibnamefont{Dominguez}},
  \bibinfo{author}{\bibfnamefont{L.}~\bibnamefont{Chiodo}},
  \bibinfo{author}{\bibfnamefont{E.}~\bibnamefont{Sheveleva}},
  \bibinfo{author}{\bibfnamefont{M.}~\bibnamefont{{Yazdi-Rizi}}},
  \bibinfo{author}{\bibfnamefont{C.}~\bibnamefont{Bernhard}},
  \bibinfo{author}{\bibfnamefont{A.}~\bibnamefont{Rubio}}, \bibnamefont{and}
  \bibinfo{author}{\bibfnamefont{M.}~\bibnamefont{Chergui}},
  \bibinfo{journal}{Phys Rev B} \textbf{\bibinfo{volume}{96}},
  \bibinfo{pages}{041204} (\bibinfo{year}{2017}).

\bibitem[{\citenamefont{Hybertsen and Louie}(1986)}]{PhysRevB.34.5390}
\bibinfo{author}{\bibfnamefont{M.~S.} \bibnamefont{Hybertsen}}
  \bibnamefont{and} \bibinfo{author}{\bibfnamefont{S.~G.} \bibnamefont{Louie}},
  \bibinfo{journal}{Phys. Rev. B} \textbf{\bibinfo{volume}{34}},
  \bibinfo{pages}{5390} (\bibinfo{year}{1986}).

\bibitem[{\citenamefont{Chiodo et~al.}(2010)\citenamefont{Chiodo,
  {Garc\'ia-Lastra}, Iacomino, Ossicini, Zhao, Petek, and
  Rubio}}]{chiodo_self-energy_2010}
\bibinfo{author}{\bibfnamefont{L.}~\bibnamefont{Chiodo}},
  \bibinfo{author}{\bibfnamefont{J.~M.} \bibnamefont{{Garc\'ia-Lastra}}},
  \bibinfo{author}{\bibfnamefont{A.}~\bibnamefont{Iacomino}},
  \bibinfo{author}{\bibfnamefont{S.}~\bibnamefont{Ossicini}},
  \bibinfo{author}{\bibfnamefont{J.}~\bibnamefont{Zhao}},
  \bibinfo{author}{\bibfnamefont{H.}~\bibnamefont{Petek}}, \bibnamefont{and}
  \bibinfo{author}{\bibfnamefont{A.}~\bibnamefont{Rubio}},
  \bibinfo{journal}{Phys. Rev. B} \textbf{\bibinfo{volume}{82}},
  \bibinfo{pages}{045207} (\bibinfo{year}{2010}).

\bibitem[{\citenamefont{Rangel et~al.}(2018)\citenamefont{Rangel, {Del Ben},
  Varsano, Antonius, Bruneval, {da Jornada}, {van Setten}, Orhan, {O'Regan},
  Canning et~al.}}]{rang+18tobe}
\bibinfo{author}{\bibfnamefont{T.}~\bibnamefont{Rangel}},
  \bibinfo{author}{\bibfnamefont{M.}~\bibnamefont{{Del Ben}}},
  \bibinfo{author}{\bibfnamefont{D.}~\bibnamefont{Varsano}},
  \bibinfo{author}{\bibfnamefont{G.}~\bibnamefont{Antonius}},
  \bibinfo{author}{\bibfnamefont{F.}~\bibnamefont{Bruneval}},
  \bibinfo{author}{\bibfnamefont{F.}~\bibnamefont{{da Jornada}}},
  \bibinfo{author}{\bibfnamefont{M.}~\bibnamefont{{van Setten}}},
  \bibinfo{author}{\bibfnamefont{O.}~\bibnamefont{Orhan}},
  \bibinfo{author}{\bibfnamefont{D.~D.} \bibnamefont{{O'Regan}}},
  \bibinfo{author}{\bibfnamefont{A.}~\bibnamefont{Canning}},
  \bibnamefont{et~al.}, \bibinfo{journal}{preprint}  (\bibinfo{year}{2018}).

\bibitem[{\citenamefont{Gonze et~al.}(2009)\citenamefont{Gonze, Amadon,
  Anglade, Beuken, Bottin, Boulanger, Bruneval, Caliste, Caracas, C\^ot\'e
  et~al.}}]{Gonze2009}
\bibinfo{author}{\bibfnamefont{X.}~\bibnamefont{Gonze}},
  \bibinfo{author}{\bibfnamefont{B.}~\bibnamefont{Amadon}},
  \bibinfo{author}{\bibfnamefont{P.-M.} \bibnamefont{Anglade}},
  \bibinfo{author}{\bibfnamefont{J.-M.} \bibnamefont{Beuken}},
  \bibinfo{author}{\bibfnamefont{F.}~\bibnamefont{Bottin}},
  \bibinfo{author}{\bibfnamefont{P.}~\bibnamefont{Boulanger}},
  \bibinfo{author}{\bibfnamefont{F.}~\bibnamefont{Bruneval}},
  \bibinfo{author}{\bibfnamefont{D.}~\bibnamefont{Caliste}},
  \bibinfo{author}{\bibfnamefont{R.}~\bibnamefont{Caracas}},
  \bibinfo{author}{\bibfnamefont{M.}~\bibnamefont{C\^ot\'e}},
  \bibnamefont{et~al.}, \bibinfo{journal}{Comput. Phys. Commun.}
  \textbf{\bibinfo{volume}{180}}, \bibinfo{pages}{2582 }
  (\bibinfo{year}{2009}).

\bibitem[{\citenamefont{Deslippe et~al.}(2012)\citenamefont{Deslippe,
  Samsonidze, Strubbe, Jain, Cohen, and Louie}}]{Deslippe2012}
\bibinfo{author}{\bibfnamefont{J.}~\bibnamefont{Deslippe}},
  \bibinfo{author}{\bibfnamefont{G.}~\bibnamefont{Samsonidze}},
  \bibinfo{author}{\bibfnamefont{D.~A.} \bibnamefont{Strubbe}},
  \bibinfo{author}{\bibfnamefont{M.}~\bibnamefont{Jain}},
  \bibinfo{author}{\bibfnamefont{M.~L.} \bibnamefont{Cohen}}, \bibnamefont{and}
  \bibinfo{author}{\bibfnamefont{S.~G.} \bibnamefont{Louie}},
  \bibinfo{journal}{Comput. Phys. Commun.} \textbf{\bibinfo{volume}{183}},
  \bibinfo{pages}{1269 } (\bibinfo{year}{2012}).

\bibitem[{\citenamefont{Hupfer et~al.}(2017{\natexlab{b}})\citenamefont{Hupfer,
  Vines, Monakhov, Svensson, and Herklotz}}]{PhysRevB.96.085203}
\bibinfo{author}{\bibfnamefont{A.}~\bibnamefont{Hupfer}},
  \bibinfo{author}{\bibfnamefont{L.}~\bibnamefont{Vines}},
  \bibinfo{author}{\bibfnamefont{E.~V.} \bibnamefont{Monakhov}},
  \bibinfo{author}{\bibfnamefont{B.~G.} \bibnamefont{Svensson}},
  \bibnamefont{and} \bibinfo{author}{\bibfnamefont{F.}~\bibnamefont{Herklotz}},
  \bibinfo{journal}{Phys Rev B} \textbf{\bibinfo{volume}{96}},
  \bibinfo{pages}{085203} (\bibinfo{year}{2017}{\natexlab{b}}).

\bibitem[{\citenamefont{Filippone et~al.}(2009)\citenamefont{Filippone,
  Mattioli, Alippi, and Amore~Bonapasta}}]{filippone_properties_2009}
\bibinfo{author}{\bibfnamefont{F.}~\bibnamefont{Filippone}},
  \bibinfo{author}{\bibfnamefont{G.}~\bibnamefont{Mattioli}},
  \bibinfo{author}{\bibfnamefont{P.}~\bibnamefont{Alippi}}, \bibnamefont{and}
  \bibinfo{author}{\bibfnamefont{A.}~\bibnamefont{Amore~Bonapasta}},
  \bibinfo{journal}{Phys. Rev. B} \textbf{\bibinfo{volume}{80}},
  \bibinfo{pages}{245203} (\bibinfo{year}{2009}).

\bibitem[{\citenamefont{{Stausholm-M\o{}ller}
  et~al.}(2010)\citenamefont{{Stausholm-M\o{}ller}, Kristoffersen, Hinnemann,
  Madsen, and Hammer}}]{stausholm-moller_dft+u_2010}
\bibinfo{author}{\bibfnamefont{J.}~\bibnamefont{{Stausholm-M\o{}ller}}},
  \bibinfo{author}{\bibfnamefont{H.~H.} \bibnamefont{Kristoffersen}},
  \bibinfo{author}{\bibfnamefont{B.}~\bibnamefont{Hinnemann}},
  \bibinfo{author}{\bibfnamefont{G.~K.~H.} \bibnamefont{Madsen}},
  \bibnamefont{and} \bibinfo{author}{\bibfnamefont{B.}~\bibnamefont{Hammer}},
  \bibinfo{journal}{J. Chem. Phys.} \textbf{\bibinfo{volume}{133}},
  \bibinfo{pages}{144708} (\bibinfo{year}{2010}).

\bibitem[{\citenamefont{De\'ak et~al.}(2011)\citenamefont{De\'ak, Aradi, and
  Frauenheim}}]{deak_polaronic_2011}
\bibinfo{author}{\bibfnamefont{P.}~\bibnamefont{De\'ak}},
  \bibinfo{author}{\bibfnamefont{B.}~\bibnamefont{Aradi}}, \bibnamefont{and}
  \bibinfo{author}{\bibfnamefont{T.}~\bibnamefont{Frauenheim}},
  \bibinfo{journal}{Phys. Rev. B} \textbf{\bibinfo{volume}{83}},
  \bibinfo{pages}{155207} (\bibinfo{year}{2011}).

\bibitem[{\citenamefont{Chen et~al.}(2018)\citenamefont{Chen, Hao, Jin, Wei,
  Feng, Jia, Yi, Rohlfing, Liu, and Ma}}]{chen2018origin}
\bibinfo{author}{\bibfnamefont{T.}~\bibnamefont{Chen}},
  \bibinfo{author}{\bibfnamefont{Y.-n.} \bibnamefont{Hao}},
  \bibinfo{author}{\bibfnamefont{F.}~\bibnamefont{Jin}},
  \bibinfo{author}{\bibfnamefont{M.}~\bibnamefont{Wei}},
  \bibinfo{author}{\bibfnamefont{J.}~\bibnamefont{Feng}},
  \bibinfo{author}{\bibfnamefont{R.}~\bibnamefont{Jia}},
  \bibinfo{author}{\bibfnamefont{Z.}~\bibnamefont{Yi}},
  \bibinfo{author}{\bibfnamefont{M.}~\bibnamefont{Rohlfing}},
  \bibinfo{author}{\bibfnamefont{C.}~\bibnamefont{Liu}}, \bibnamefont{and}
  \bibinfo{author}{\bibfnamefont{Y.}~\bibnamefont{Ma}},
  \bibinfo{journal}{Physical Review B} \textbf{\bibinfo{volume}{98}},
  \bibinfo{pages}{205135} (\bibinfo{year}{2018}).

\bibitem[{\citenamefont{Pang et~al.}(2013)\citenamefont{Pang, Lindsay, and
  Thornton}}]{pang2013structure}
\bibinfo{author}{\bibfnamefont{C.~L.} \bibnamefont{Pang}},
  \bibinfo{author}{\bibfnamefont{R.}~\bibnamefont{Lindsay}}, \bibnamefont{and}
  \bibinfo{author}{\bibfnamefont{G.}~\bibnamefont{Thornton}},
  \bibinfo{journal}{Chemical reviews} \textbf{\bibinfo{volume}{113}},
  \bibinfo{pages}{3887} (\bibinfo{year}{2013}).

\bibitem[{\citenamefont{Yin et~al.}(2018)\citenamefont{Yin, Wen, Zhou, Selloni,
  and Liu}}]{yin2018excess}
\bibinfo{author}{\bibfnamefont{W.-J.} \bibnamefont{Yin}},
  \bibinfo{author}{\bibfnamefont{B.}~\bibnamefont{Wen}},
  \bibinfo{author}{\bibfnamefont{C.}~\bibnamefont{Zhou}},
  \bibinfo{author}{\bibfnamefont{A.}~\bibnamefont{Selloni}}, \bibnamefont{and}
  \bibinfo{author}{\bibfnamefont{L.-M.} \bibnamefont{Liu}},
  \bibinfo{journal}{Surface Science Reports} \textbf{\bibinfo{volume}{73}},
  \bibinfo{pages}{58} (\bibinfo{year}{2018}).

\bibitem[{\citenamefont{Landmann et~al.}(2012)\citenamefont{Landmann, Rauls,
  and Schmidt}}]{landmann_electronic_2012}
\bibinfo{author}{\bibfnamefont{M.}~\bibnamefont{Landmann}},
  \bibinfo{author}{\bibfnamefont{E.}~\bibnamefont{Rauls}}, \bibnamefont{and}
  \bibinfo{author}{\bibfnamefont{W.~G.} \bibnamefont{Schmidt}},
  \bibinfo{journal}{J. Phys. Condens. Matter} \textbf{\bibinfo{volume}{24}},
  \bibinfo{pages}{195503} (\bibinfo{year}{2012}).

\bibitem[{\citenamefont{Pizzi et~al.}(2016)\citenamefont{Pizzi, Cepellotti,
  Sabatini, Marzari, and Kozinsky}}]{PIZZI2016218}
\bibinfo{author}{\bibfnamefont{G.}~\bibnamefont{Pizzi}},
  \bibinfo{author}{\bibfnamefont{A.}~\bibnamefont{Cepellotti}},
  \bibinfo{author}{\bibfnamefont{R.}~\bibnamefont{Sabatini}},
  \bibinfo{author}{\bibfnamefont{N.}~\bibnamefont{Marzari}}, \bibnamefont{and}
  \bibinfo{author}{\bibfnamefont{B.}~\bibnamefont{Kozinsky}},
  \bibinfo{journal}{Comput. Mater. Sci.} \textbf{\bibinfo{volume}{111}},
  \bibinfo{pages}{218} (\bibinfo{year}{2016}).

\end{thebibliography}

\end{document}